\documentclass[sigconf,10pt]{acmart}
\setcopyright{acmcopyright}
\copyrightyear{2021}
\acmYear{2021}
\acmDOI{XX.XXXX/XXXXXXX}

\acmConference[WiNTECH '21]{WiNTECH '21: ACM Workshop on Wireless Network Testbeds, Experimental evaluation & Characterization}{October 29, 2021}{Virtual Conference}
\acmBooktitle{WiNTECH '21: ACM Workshop on Wireless Network Testbeds, Experimental evaluation \\downarrow \& Characterization,
  October 29, 2021, Virtual Conference}
\acmPrice{15.00}
\acmISBN{978-1-4503-XXXX-X/18/06}

\usepackage{color,graphicx}
\usepackage{hyperref}
\usepackage{amsmath}
\usepackage{graphicx}
\usepackage{caption}
\usepackage{subcaption}
\usepackage{multirow}
\usepackage[shortlabels]{enumitem}
\usepackage{epstopdf}
\usepackage{epsfig}
\definecolor{Gray}{gray}{0.9}
\usepackage{tikz}

\PassOptionsToPackage{linktocpage}{hyperref}

\makeatletter

\newcommand{\Rmnum}[1]{\expandafter\@slowromancap\romannumeral #1@}

\makeatother
\begin{document}
\title{A Comparison Study of Cellular Deployments in Chicago and Miami Using Apps on Smartphones}
\author{Muhammad Iqbal Rochman}
\affiliation{%
  \institution{Univ. of Chicago}}
\email{muhiqbalcr@uchicago.edu}

\author{Vanlin Sathya}
\affiliation{%
  \institution{Univ. of Chicago}
}
\email{vanlin@uchicago.edu}

\author{Norlen Nunez}
\affiliation{%
  \institution{FIU, Miami}
}
\email{nnune047@fiu.edu}

\author{Damian Fernandez}
\affiliation{%
  \institution{FIU, Miami}
}
\email{dfern265@fiu.edu}

\author{Monisha Ghosh}
\affiliation{%
  \institution{Univ. of Chicago}
}
\email{monisha@uchicago.edu}

\author{Ahmed S. Ibrahim}
\affiliation{%
  \institution{FIU, Miami}
}
\email{aibrahim@fiu.edu}

\author{William Payne}
\affiliation{%
  \institution{Univ. of Chicago}
}
\email{billpayne@uchicago.edu}

\begin{abstract}
Cellular operators have begun deploying 5G New Radio (NR) in all available bands: low (< 1 GHz), mid (1 - 6 GHz), and high (> 24 GHz) to exploit the different capabilities of each. At the same time, traditional 4G Long Term Evolution (LTE) deployments are being enhanced with the addition of bands in the unlicensed 5 GHz (using License Assisted Access, or LAA) and the 3.5 GHz Citizens Broadband Radio Service (CBRS) resulting in throughput performance comparable to 5G in mid-band. We present a detailed study comparing 4G and 5G deployments, in all bands in Chicago, and focused mmWave measurements and analysis in Miami. Our methodology, based on commercial and custom apps, is scalable for crowdsourcing measurements on a large scale and provides detailed data (throughput, latency, signal strength, etc.) on actual deployments. Our main conclusions based on the measurements are (i) optimized 4G networks in mid-band are comparable in both throughput and latency to current deployments of 5G (both standalone (SA) and non-standalone (NSA)) and (ii) mmWave 5G, even in NSA mode, can deliver multi-Gbps throughput reliably if the installation is dense enough, but performance is still brittle due to the propagation limitations imposed by distance and body-loss. Thus, while 5G demonstrates significant early promise, further work needs to be done to ensure that the stated goals of 5G are met.

\end{abstract}


\maketitle
\pagestyle{plain}
\section{Introduction}
In response to high-throughput and low-latency requirements of emerging applications, cellular network operators are aggressively rolling out 5G New Radio (5G NR), as specified by 3GPP in Release 15~\footnote{https://www.3gpp.org/release-15}, in Frequency Range 1 (FR1) which includes low-band (< 1 GHz) and mid-band (1 - 6 GHz) frequencies, and Frequency Range 2 (FR2) which includes the latest high-band frequencies in the mmWave range (> 24 GHz). 3GPP specifies two deployment modes for 5G: Non-Standalone (NSA), requiring a 4G primary channel and Standalone (SA), without that requirement. Presently, most 5G deployments in the US are NSA, but SA is beginning to be deployed in limited areas as well.

\begin{figure*}[h!]
\begin{subfigure}{.24\textwidth}
  \centering
 \includegraphics[width=3.8cm, height=6cm]{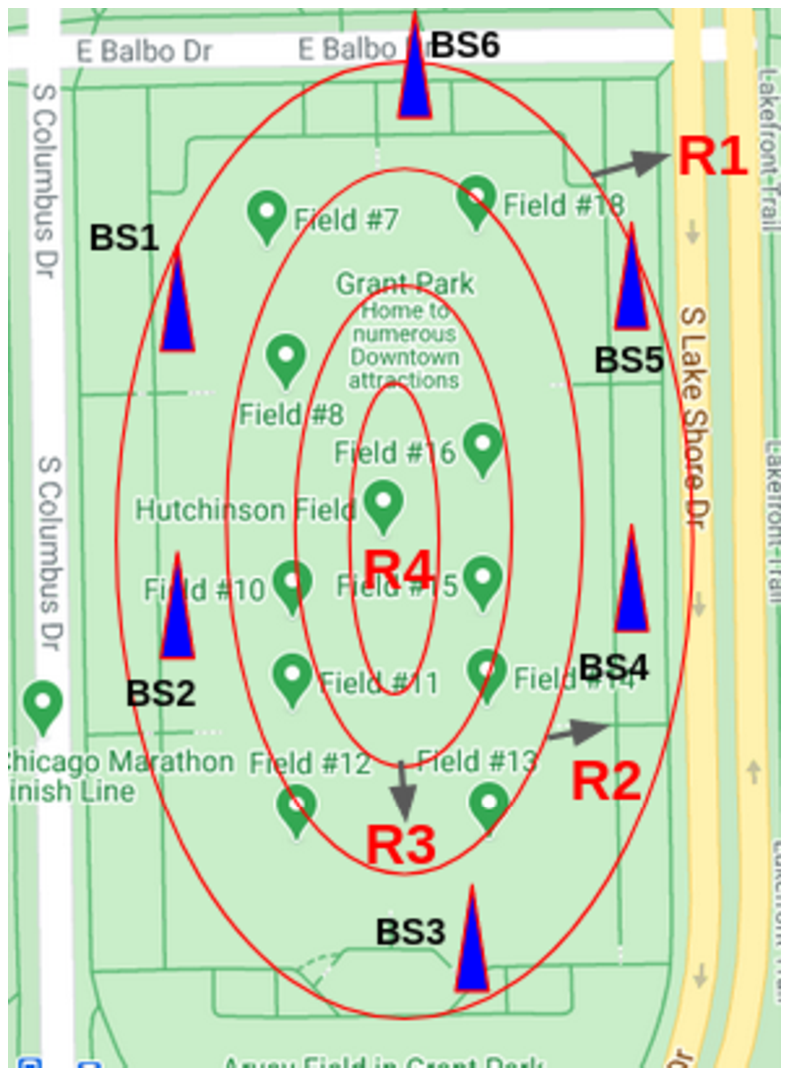}
    \caption{Experiment Location}\label{fig:gpMap}
      \end{subfigure}
 \begin{subfigure}{.24\textwidth}
  \centering
  \includegraphics[width=3.5cm, height=6cm]{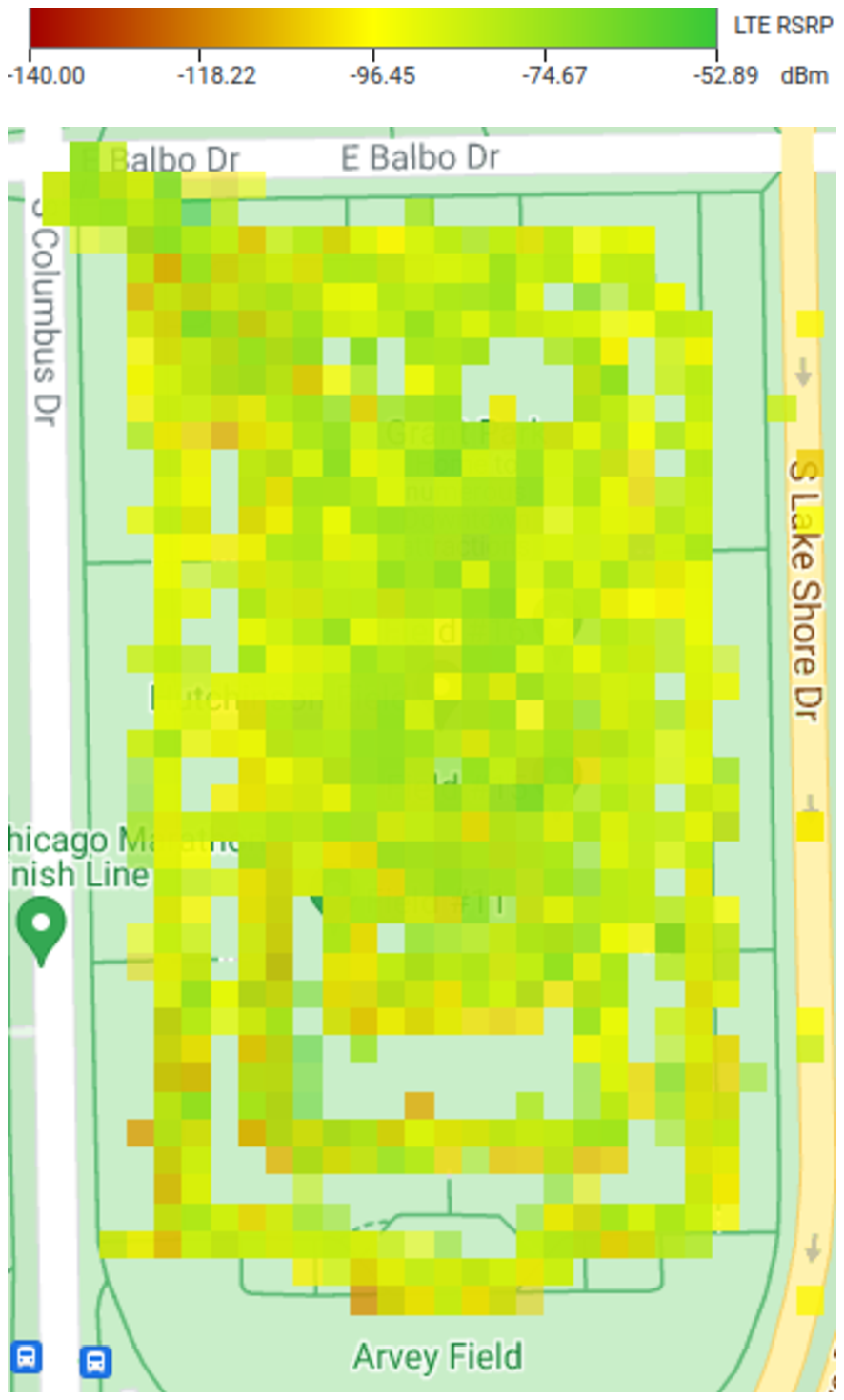}
  \caption{4G Licensed}\label{fig:heatmap4G}
       \end{subfigure}
           \begin{subfigure}{.24\textwidth}
  \centering 
  \includegraphics[width=3.5cm, height=6cm]{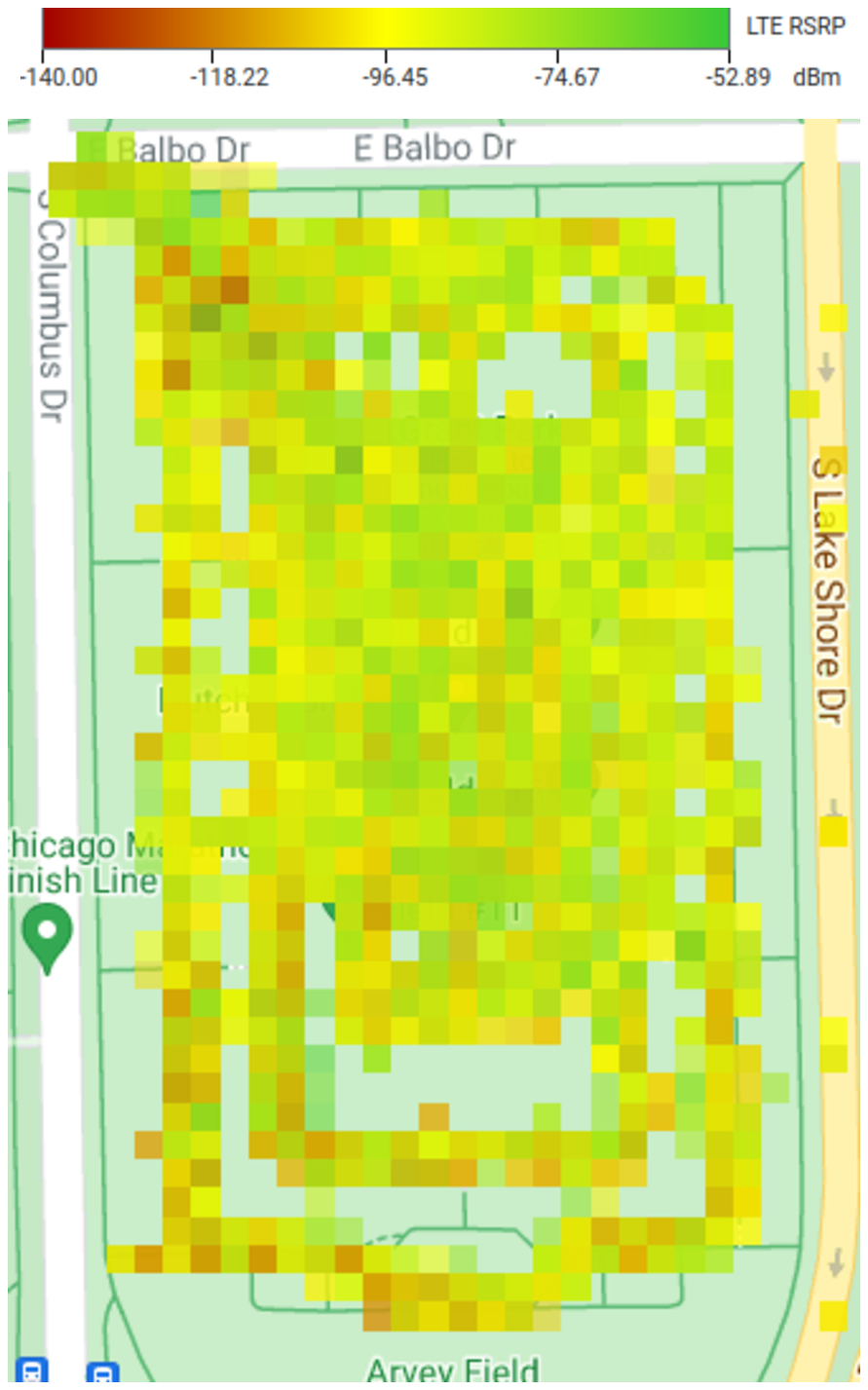}
  \caption{4G Lic., LAA and CBRS}\label{fig:heatmap4Gp}
      \end{subfigure}
 \begin{subfigure}{.24\textwidth}
  \centering 
     \includegraphics[width=3.5cm, height=6cm]{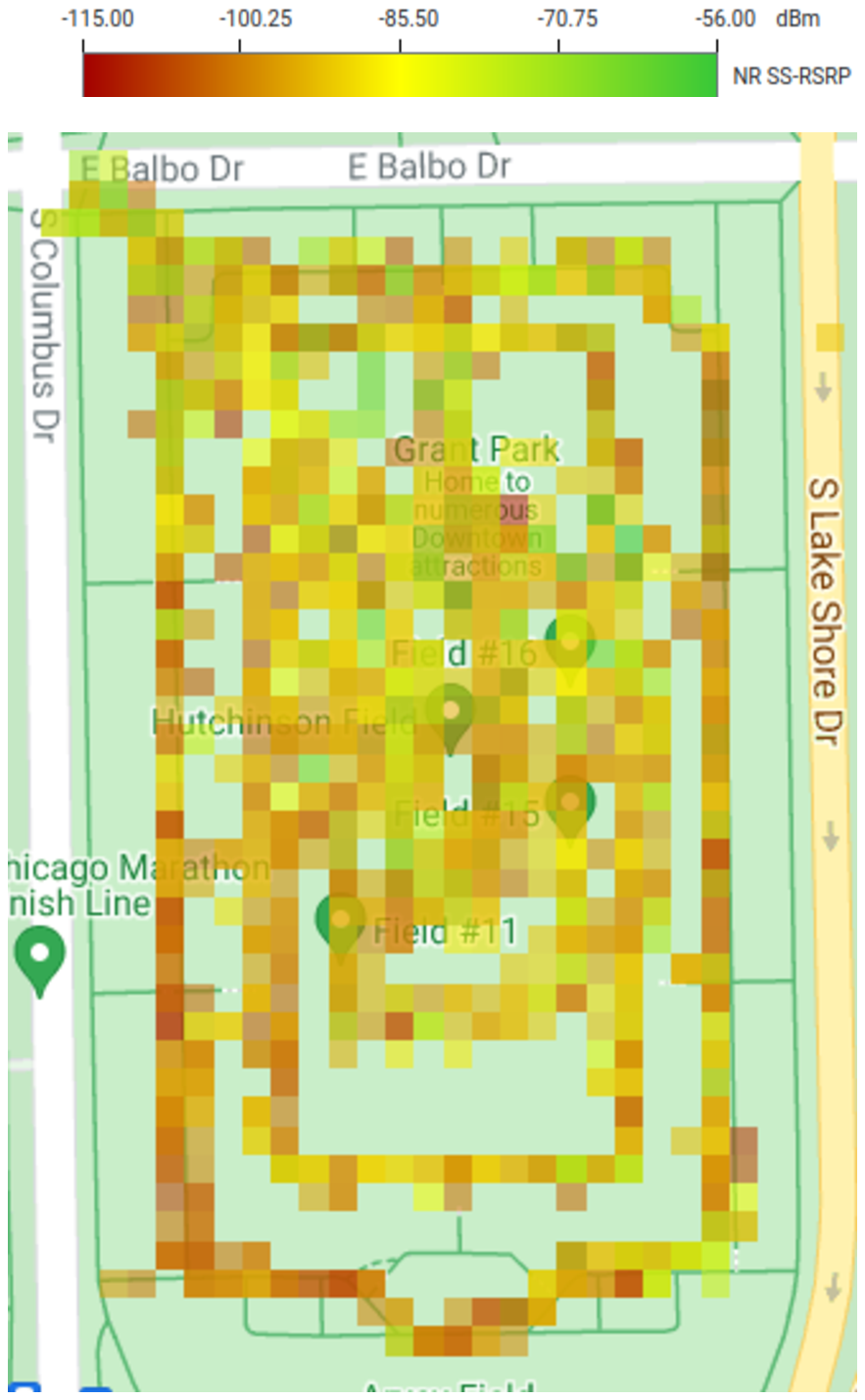}
       \caption{5G FR1 and FR2}\label{fig:heatmap5G}
      \end{subfigure} 
      \caption{Hutchinson Field Overview: Verizon, T-Mobile and AT\&T 4G, 4G+LAA/CBRS, and 5G Coverage}
\end{figure*}

\begin{table}
\small
\caption{Operator Deployment in Hutchinson Field, Chicago, and Downtown Miami (TDD bands in bold).}
\vspace{-0.2cm}
\begin{tabular}{|p{1.2cm}|p{1.39cm}|p{1.0cm}|p{1.0cm}|p{2.5cm}|}
\hline
\textbf{Operator} & \textbf{Deploy-ment} & \textbf{5G Freq.} & \textbf{5G Op. Bands} & \textbf{4G Op. Bands (LAA:46,CBRS:48)} \\
\hline
\hline
Verizon & 4G+LAA \& CBRS, 5G & Low, High & n5, \textbf{n260} &  2, 4, 5, 13, \textbf{46}, \textbf{48}, 66 \\
\hline
T-Mobile & 4G, 5G & Low, Mid & \textbf{n41}, n71 & 2, 4, 7, 12, 66 \\
\hline
AT\&T & 4G+LAA & Low & n5 & 2, 4, 12, 14, 30, 46, 66 \\
\hline
Verizon (Miami)  & 4G+LAA, 5G & High & \textbf{n261} &  2, 4, 13, \textbf{46}, 66 \\
\hline

\end{tabular}
\label{opd}
\vspace{-0.4cm}
\end{table}

5G mmWave has the potential to deliver up to 2 Gbps downlink throughput, verified by our measurements, and possibly higher as device and network performance improves. However, this is limited to outdoor deployments and is more susceptible to degradation due to blockage from the body and other objects in the environment.
At the lower end of the spectrum, 5G can provide extended, robust coverage but with lower throughput due to the limited available bandwidth\footnote{At the time of writing, most 5G NSA phones, including the Google Pixel 5, can only aggregate one 5G channel in FR1, while multiple channels can be aggregated in FR2.}.
Additionally, the current 5G NSA deployment may increase latency due to the overhead imposed by dual connectivity (DC) where the primary 4G channel and the secondary 5G channel may be transmitted from base-stations (BSs) that are not co-located.
At the same time, there are increasing deployments of 4G in the mid-band using unlicensed 5 GHz (with License Assisted Access (LAA)) and 3.55 - 3.7 GHz Citizen Band Radio Service (CBRS) bands as secondary aggregated channels.
While LAA deployments have been widespread over the last couple of years, CBRS deployments have started appearing just recently since the completion of the CBRS auction in Fall 2020. We have previously studied in detail the coexistence issues between 5 GHz LAA and Wi-Fi deployments in downtown Chicago and the University of Chicago \cite{sathya2020measurement,sathya2021hidden}, where we measured an average cellular throughput of 150 Mbps when 60 MHz in the unlicensed band (using three aggregated 20 MHz channels in 5 GHz) is used along with a primary 15 MHz - 20 MHz bandwidth primary channel: this is 6$\times$ the average throughput of the licensed primary band alone.

Thus, cellular deployments today have become increasingly complex, with a plethora of technologies and aggregated bands. These are extremely difficult to replicate for research purposes, even in large-scale test-beds such as NSF's Platforms for Advanced Wireless Research (PAWR)~\cite{PAWR} and others. Hence, in this paper, we develop a scalable methodology for collecting cellular network measurements and present a thorough analysis and comparison of the various deployment scenarios we observed in 2 major cities: Chicago and Miami, to inform researchers of the nature of problems that arise in actual cellular deployments today. Since these are not experimental platforms where innovative ideas can be tested, our objective\footnote{This work was supported by NSF under grant CNS-1618836.} in performing these measurements is to use these results to uncover problems that can then be investigated in detail in the test-beds. \\

\textbf{Related Work}: There is very little work similar to ours, using complex, deployed networks for measurements and analysis instead of test-beds.  
In~\cite{narayanan2020first}, the authors conducted an extensive study of 5G mmWave performance in a dense urban environment and analyzed the hand-off mechanism in 5G and the impact on mobile performance. Authors in~\cite{narayanan2020lumos5g} identified vital device-side factors that affect 5G performance and quantified to what extent the user can predict the 5G throughput.
In~\cite{xu2020understanding}, the authors analyzed 5G mid-band and suggested that the upper-layer protocols, wireless path, computing, and radio hardware architecture need to co-evolve with 5G to form an ecosystem to unleash its potential fully.

The above literature does not discuss the effect of physical and MAC layer parameters such as Reference Signal Received Power (RSRP), Reference Signal Received Quality (RSRQ) and Resource Block (RB) allocations on throughput. Section 2 describes our methodology for extracting these and other parameters using apps on smartphones. While we have collected measurements in many different areas of Chicago, Section 3 presents detailed comparisons of the three major networks only in Hutchinson Field, an outdoor park area where there are dense cellular deployments in all the major bands and technologies described above to service the crowds that are common in the summer months when popular outdoor events are hosted. Table~\ref{opd} shows the various technologies and frequency bands deployed by each operator. Section 4 presents an in-depth study of 5G mmWave performance in Miami, focusing on quantifying the performance of 5G mmWave as a function of body blockage, distance, and the number of devices connected to the base station. Finally, conclusions and future research directions are presented in Section 5.


\begin{table}[t]
\small
\caption{Measurement Apps' Features}
\vspace{-0.2cm}
\scalebox{0.9}{
\begin{tabular}{|p{1.7cm}|p{1.8cm}|p{2.1cm}|p{2.3cm}|}
\hline
\textbf{Features} & \textbf{SigCap} &  \textbf{FCC ST} &  \textbf{NSG} \\
\hline\hline
LTE Cell Information & \textbf{All cells:} PCI, EARFCN, Band, RSRP, RSRQ, RSSI + Primary cell bandwidth & \textbf{Primary cell only:} PCI, EARFCN, Band, RSRP, RSRQ & PCI, EARFCN, Band, Bandwidth, RSRP, RSRQ, RSSI, SINR, CQI, MIMO mode, RB allocation\\
\hline
5G Cell Information & 5G-RSRP and 5G-RSRQ & 5G-RSRP and 5G-RSRQ & PCI, NR-ARFCN, Band, Bandwidth, Beam ID, 5G-RSRP, 5G-RSRQ, SINR, CQI, MIMO mode, RB allocation\\
\hline
Throughput-related metrics & No & Application-level uplink/downlink throughput, latency & Application, RLC, MAC, and PHY layer uplink/downlink throughput \\
\hline
Root access & No & No & Yes \\
\hline
\end{tabular}}
\label{table:features}
\vspace{-0.3cm}
\end{table}



\begin{table}[!t]
\small
\caption{Devices used for 4G and 5G Measurements}
\vspace{-0.2cm}
\centering
\begin{tabular}{|l|c|c|c|}
\hline
\bfseries Location & \bfseries Mobile Device  & \bfseries Network Support\\
\hline
\multirow{3}{*}{\bfseries Chicago} & 2 $\times$ Google Pixel 2 & 4G Licensed Only \\
\cline{2-3}
& 2 $\times$ Google Pixel 3 & 4G Lic., LAA, CBRS \\
\cline{2-3}
& 3 $\times$ Google Pixel 5 & 4G Lic., LAA, CBRS, 5G\\
\hline
\bfseries Miami & 2 $\times$ Google Pixel 5 & 4G Lic., LAA, CBRS, 5G\\
\hline
\end{tabular}
\label{devices}
\vspace{-0.3cm}
\end{table}

\begin{figure*}[h!]
\begin{subfigure}[t]{.33\textwidth}
  \centering
 \includegraphics[width=\textwidth]{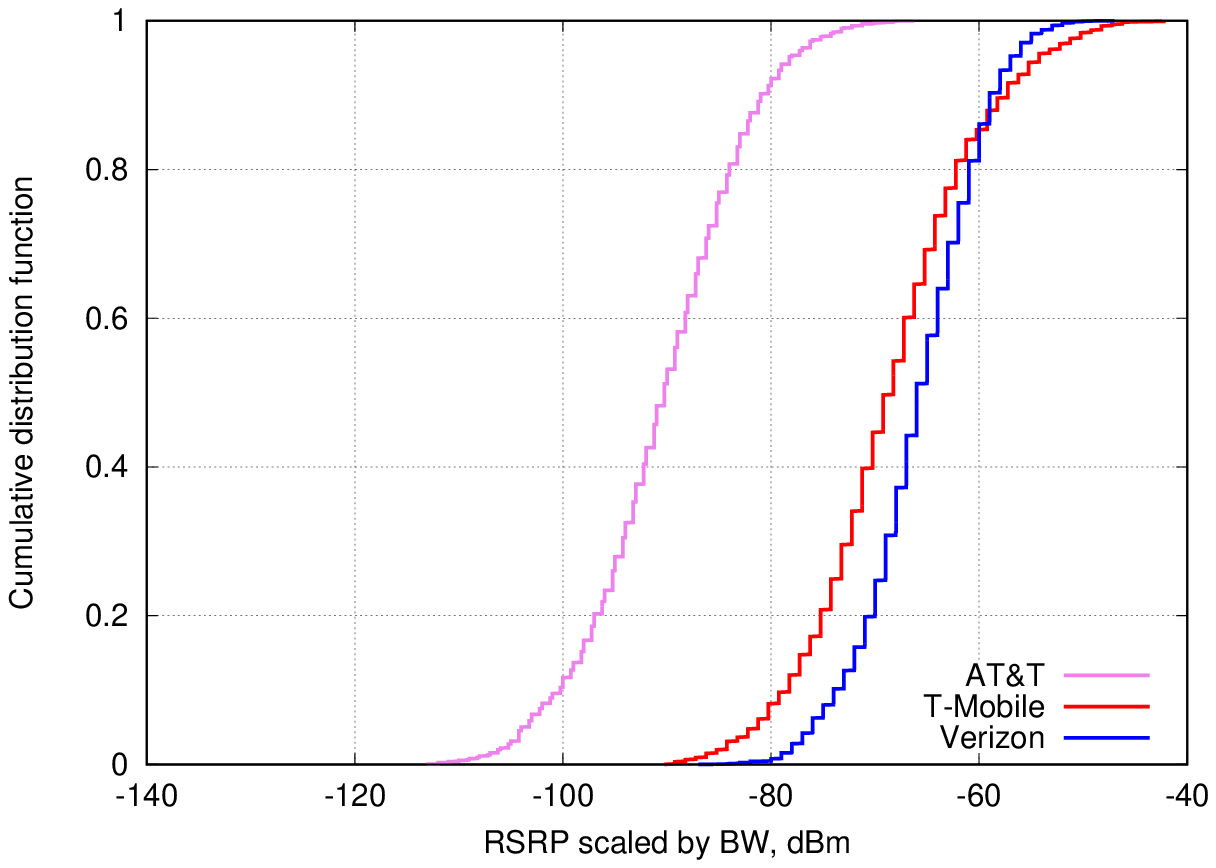}
  \caption{All: Primary BW-scaled RSRP}\label{fig:primaryRsrpBw}
\end{subfigure}
\begin{subfigure}[t]{.33\textwidth}
  \centering
 \includegraphics[width=\textwidth]{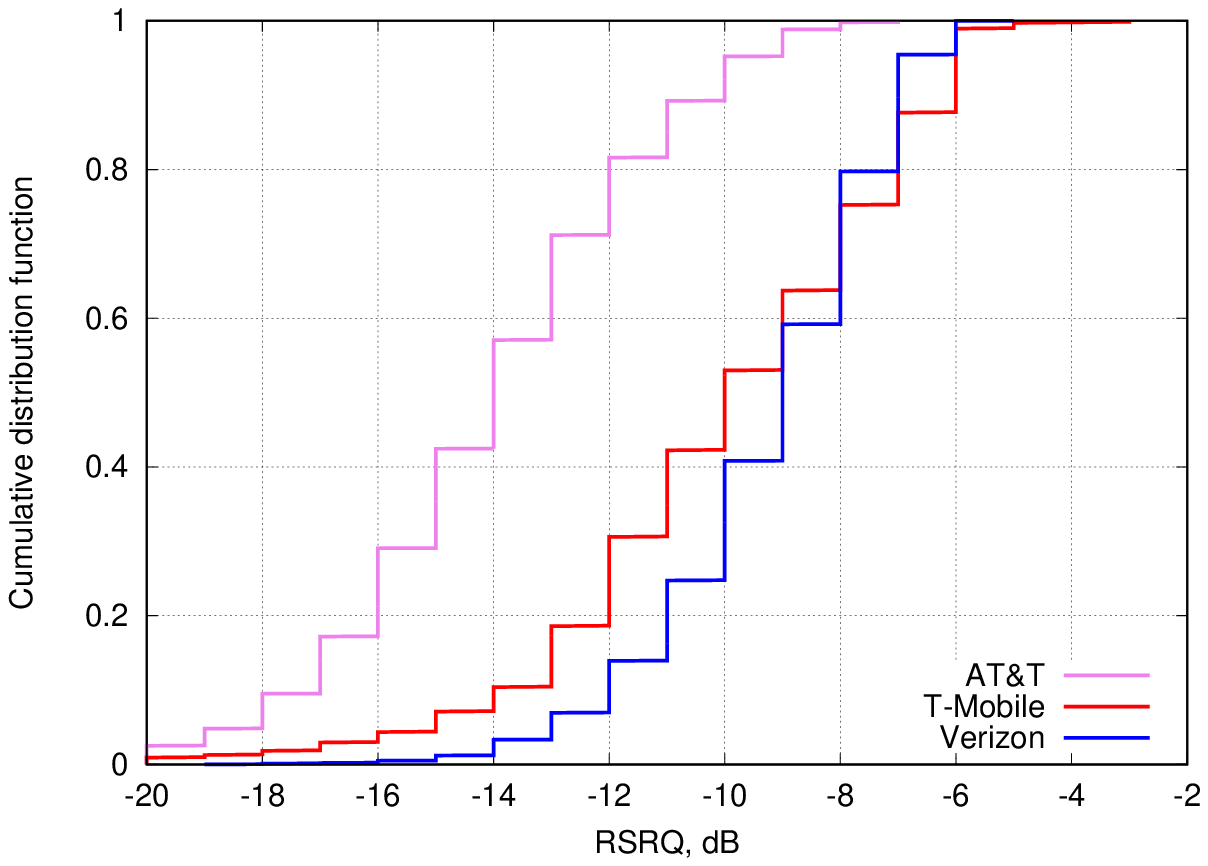}
  \caption{All: Primary RSRQ}\label{fig:primaryRsrq}
\end{subfigure}
\begin{subfigure}[t]{.33\textwidth}
  \centering
 \includegraphics[width=\textwidth]{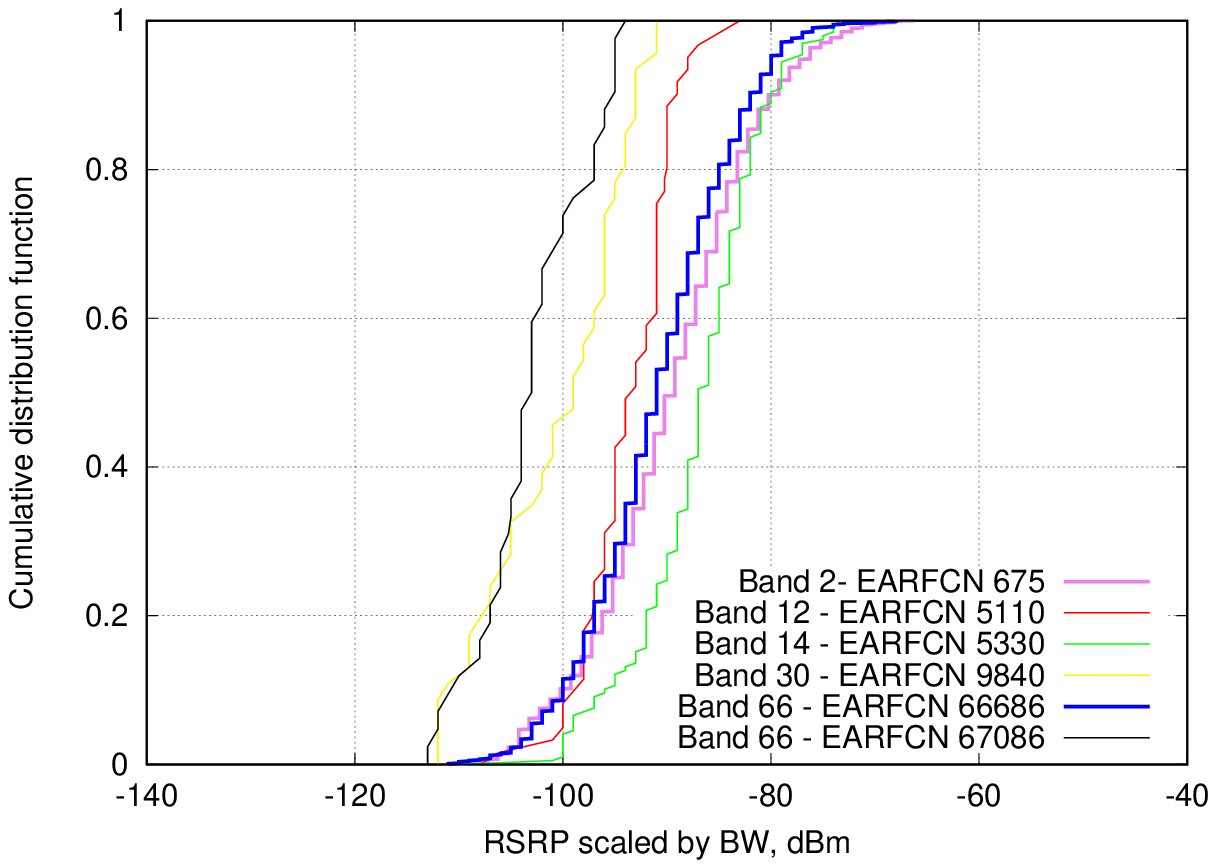}
  \caption{AT\&T: Primary BW-scaled RSRP}\label{fig:attEarfcn}
\end{subfigure}
\\
\begin{subfigure}[t]{.33\textwidth}
  \centering
 \includegraphics[width=\textwidth]{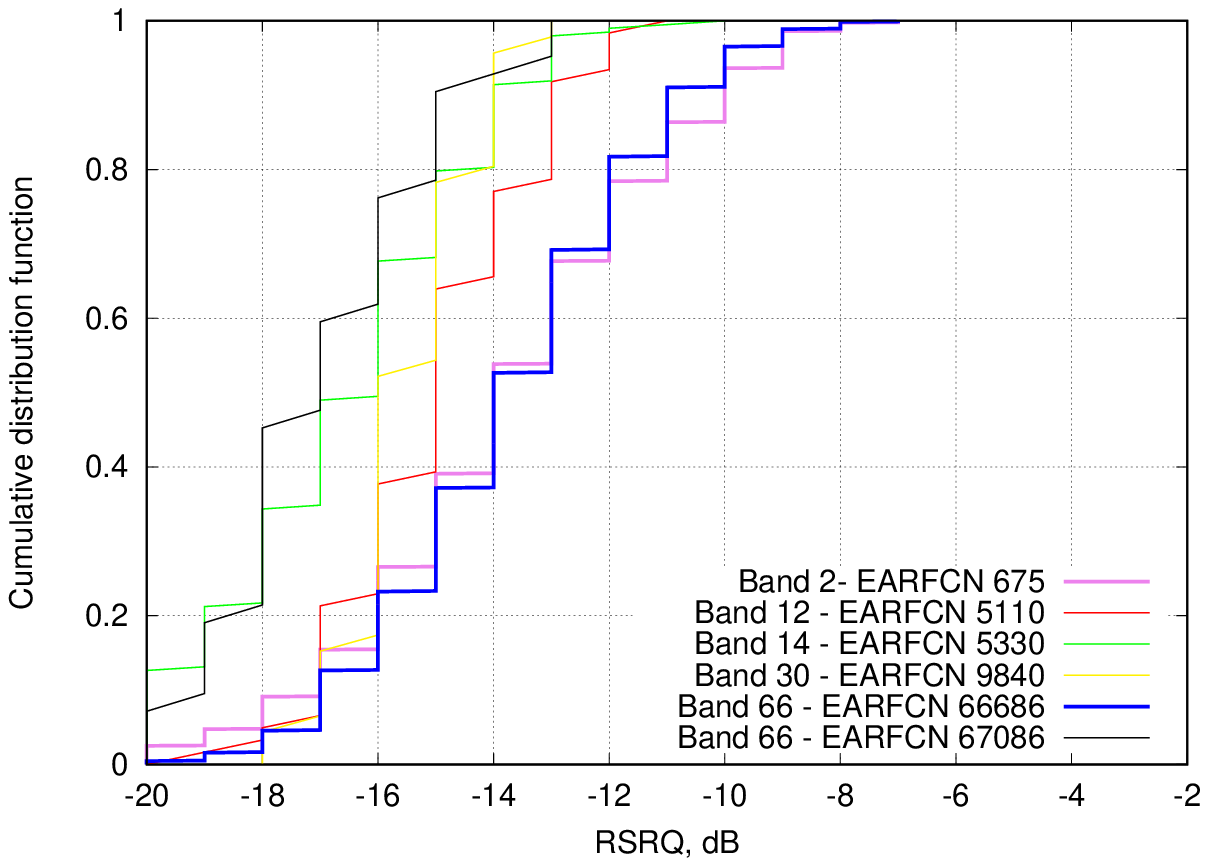}
  \caption{AT\&T: Primary RSRQ}\label{fig:attEarfcnRsrq}
\end{subfigure}
\begin{subfigure}[t]{.33\textwidth}
  \centering
    \includegraphics[width=\textwidth]{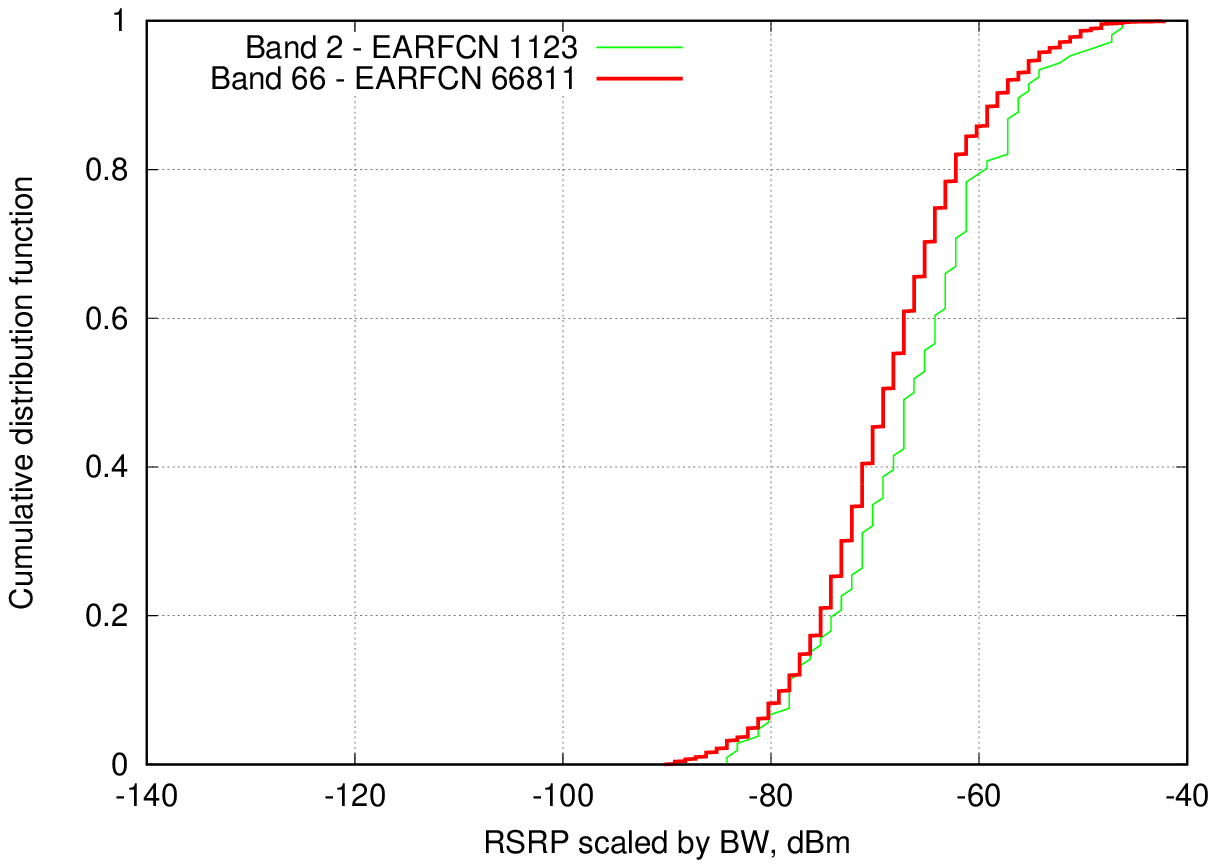}
  \caption{T-Mobile: Primary BW-scaled RSRP}\label{fig:tmoEarfcn}
\end{subfigure}
\begin{subfigure}[t]{.33\textwidth}
  \centering
    \includegraphics[width=\textwidth]{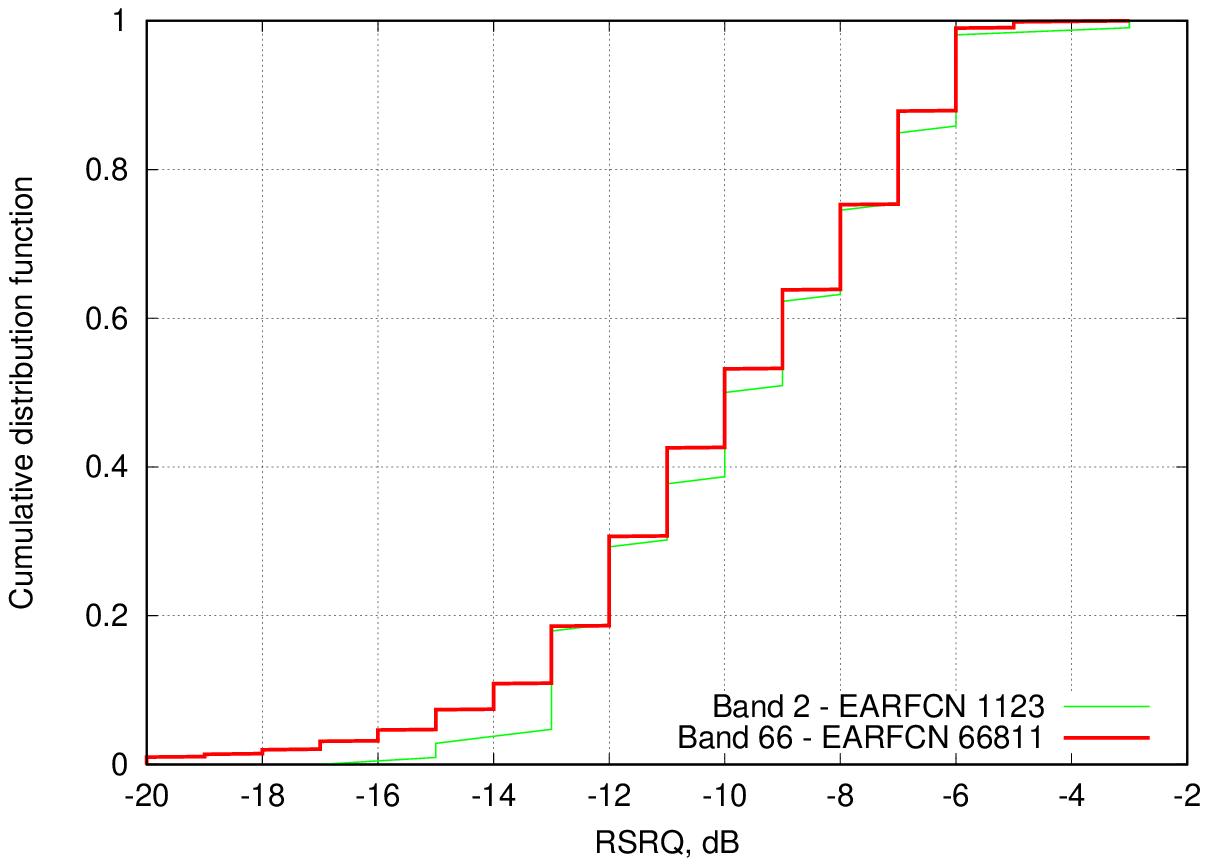}
  \caption{T-Mobile: Primary RSRQ}\label{fig:tmoEarfcnRsrq}
\end{subfigure}
\\
\begin{subfigure}[t]{.33\textwidth}
  \centering
   \includegraphics[width=\textwidth]{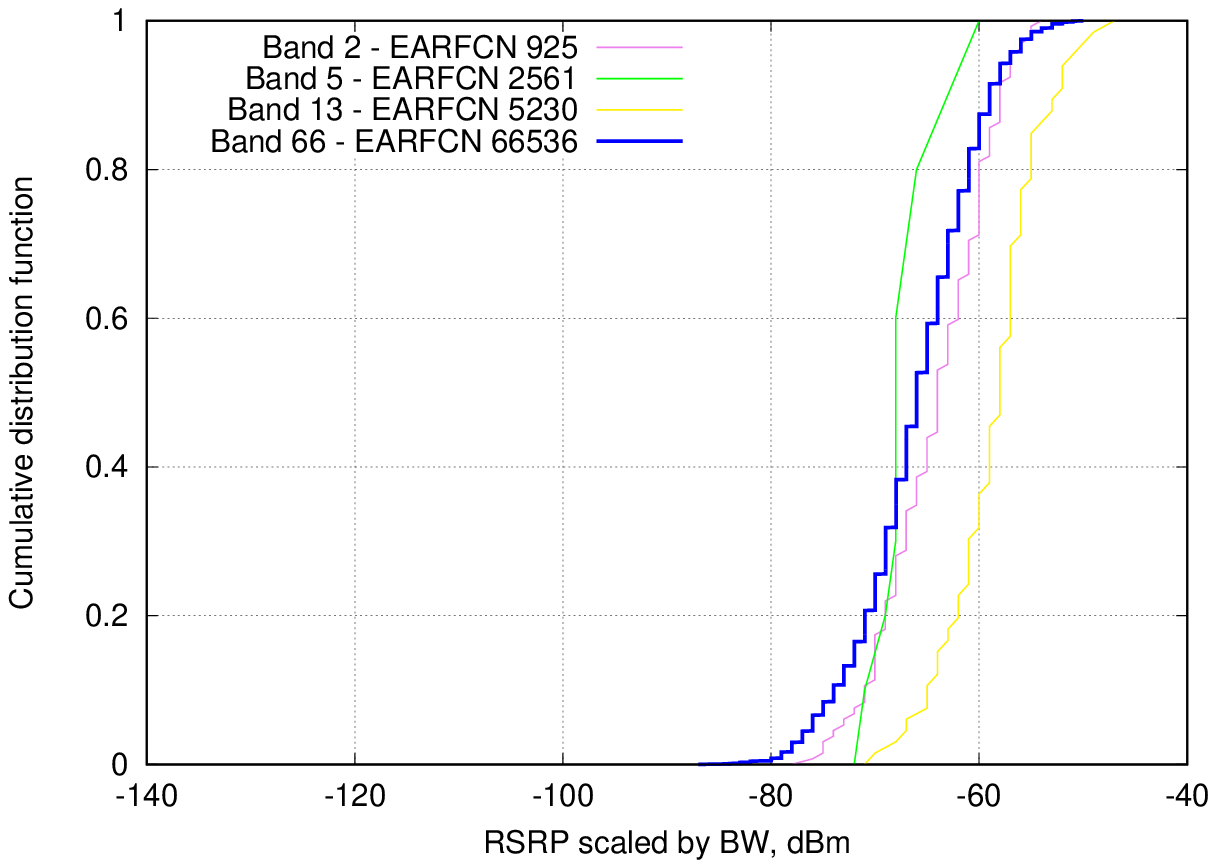}
  \caption{Verizon: Primary BW-scaled RSRP}\label{fig:vzwEarfcn}
\end{subfigure}
\begin{subfigure}[t]{.33\textwidth}
  \centering
   \includegraphics[width=\textwidth]{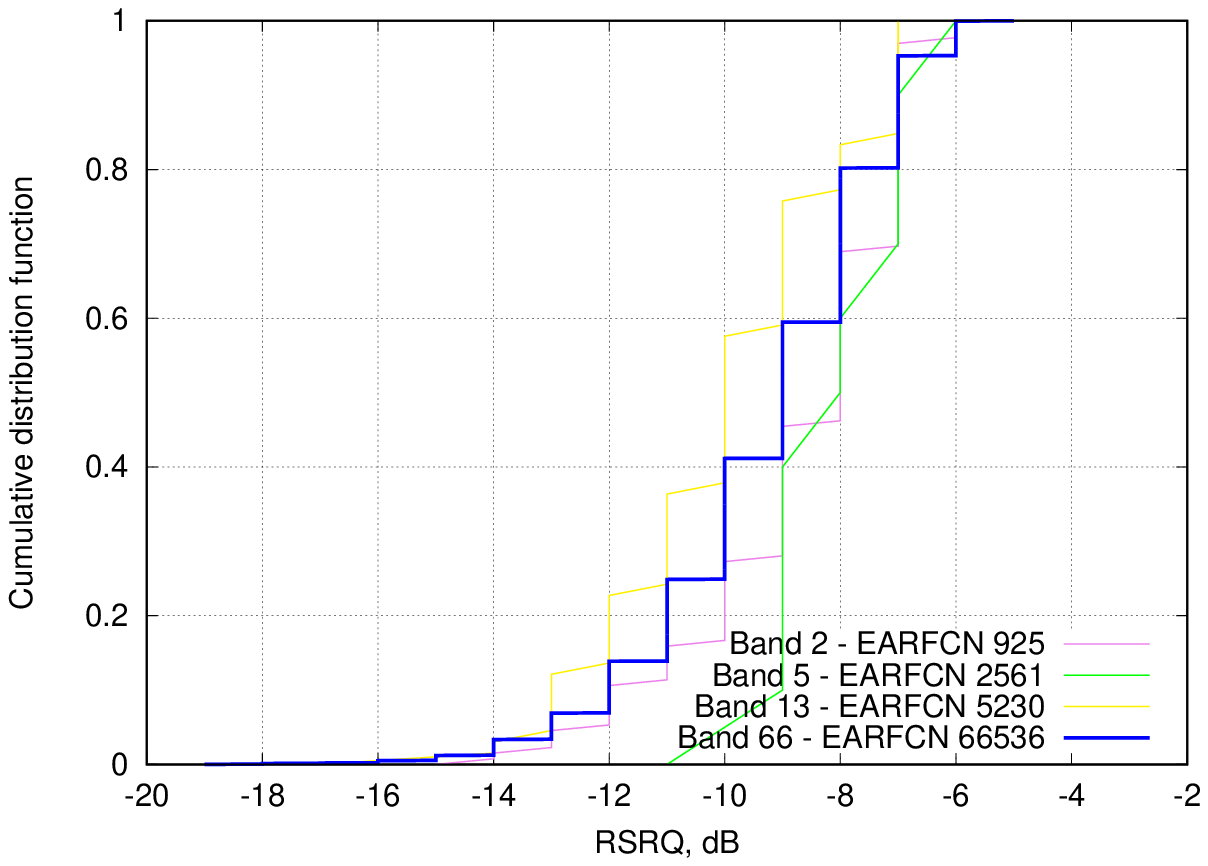}
  \caption{Verizon: Primary RSRQ}\label{fig:vzwEarfcnRsrq}
\end{subfigure}
\begin{subfigure}[t]{.33\textwidth}
  \centering
  \includegraphics[width=\textwidth]{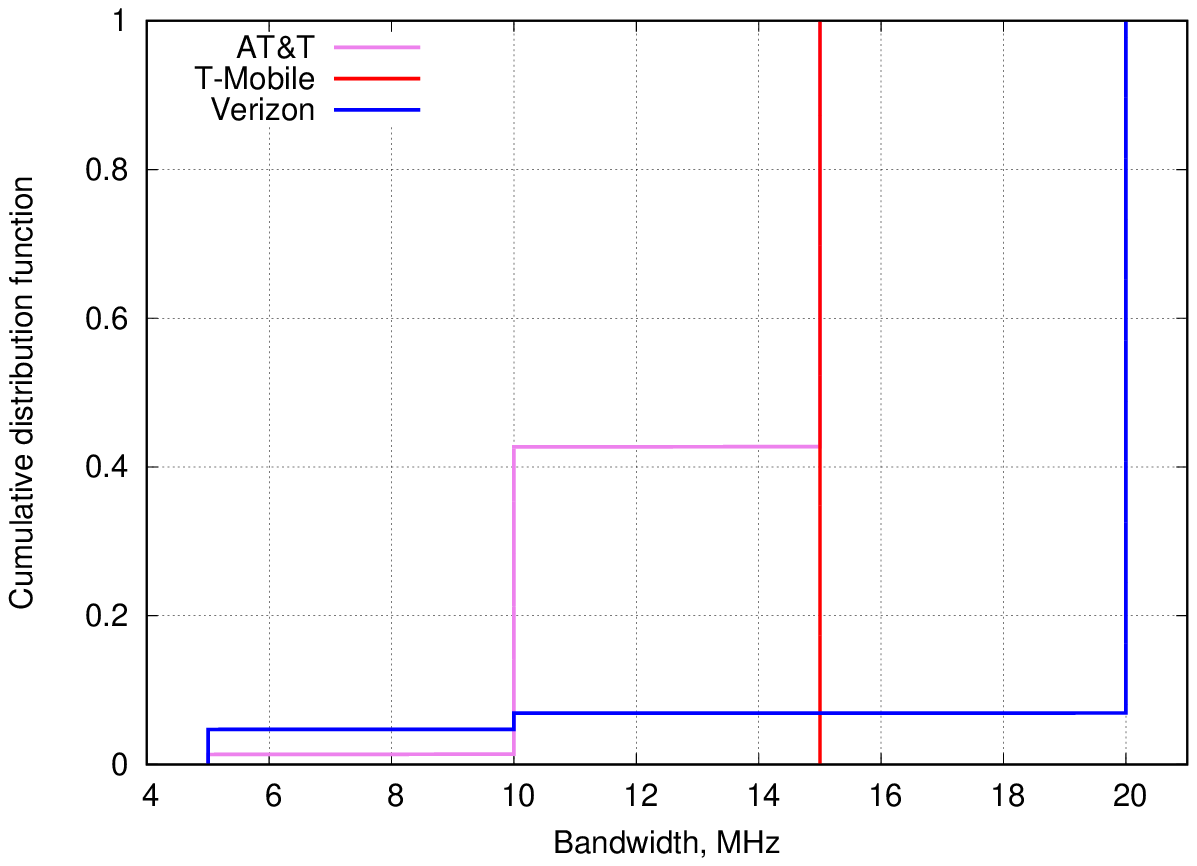}
  \caption{All: Primary BW}\label{fig:primaryBw}
\end{subfigure}
\vspace{-0.2cm}
\caption{AT\&T, T-Mobile and Verizon in Hutchinson Field: CDF of Primary Channel RSRP, RSRQ, and Bandwidth}
\label{c}
\end{figure*}

\section{Data Collection Methodology}

4G and 5G measurements were collected over several months in 2020 and 2021 in various locations in Chicago and Miami, with the intent to (i) compare the performance of the three major carriers with the greatest diversity of deployment options, and (ii) perform an in-depth study of mmWave performance. Thus, we present results from one location, Hutchinson Field, in Chicago for the former and one location in downtown Miami for the latter.
Instead of using professional drive-test equipment, smartphones equipped with apps to gather detailed network information were used: an approach that is more scalable and suitable for crowdsourcing.
This approach was also used in a recent feasibility study conducted by the Federal Communications Commission (FCC) in Colorado to map broadband availability\footnote{https://www.fcc.gov/sites/default/files/report-congress-usps-broadband-data-collection-feasibility-05242021.pdf}. The data we collected is available on our website\footnote{https://people.cs.uchicago.edu/~{}muhiqbalcr/grant-park-may-jun-2021/nr-heatmap.html}, and is available for download. 

We use three Android applications, each of which supply varying degrees of information: SigCap, developed at the University of Chicago,\footnote{https://appdistribution.firebase.google.com/pub/i/5b022e1d936d1211}, FCC Speed Test (FCC ST)\footnote{https://play.google.com/store/apps/details?id=com.samknows.fcc}, and Network Signal Guru (NSG)\footnote{https://m.qtrun.com/en/product.html}. Table~\ref{table:features} summarizes the features of these apps. 
Both SigCap and FCC ST record 4G and 5G signal information from the Android API without root access, although FCC ST only records primary 4G channel or secondary 5G channel depending on the network. We use FCC ST to collect uplink/downlink throughput and round trip latency performance, however, we omit uplink analysis due to lack of space. SigCap collects data every 10 seconds, and FCC ST every minute. 

There are some limitations imposed by the currently available APIs: (i) inability to distinguish between secondary and neighboring 4G channels, (ii) very limited 5G information, and (iii) FCC ST specifically cannot distinguish between 5G FR1 and FR2. We compensate these limitations using NSG, which uses root capability to extract information directly from the modem chipset.
However, due to the difficulty in exporting data from this app, we use NSG to study a few cases in detail and use SigCap and FCC ST for heat-maps and statistical analyses.
Using the three apps together allows us to extract detailed information about cellular network performance.

\section{Measurements in Chicago}
\subsection{Methodology and Overview}
Hutchinson Field is part of a large urban park called Grant Park in Chicago.
The area, spanning approximately 0.1 km$^2$, is shown in Fig.~\ref{fig:gpMap}.
There are dense deployments of Verizon's 4G Licensed, LAA, CBRS, and 5G as shown in Table~\ref{opd}, with fewer deployments by T-Mobile and AT\&T.
Table~\ref{devices} shows the mobile devices used for the measurements and their capabilities. Pixel 3 and Pixel 5 have root capability, required by NSG. As needed, each device is equipped with AT\&T, T-Mobile, or Verizon SIMs with unlimited data plans.
Data was collected by walking with the devices in the four different regions, with different radii, as shown in Fig.~\ref{fig:gpMap}: Outer Region Round 1 (R1), Inner Region Round 2 (R2), Inner Region Round 3 (R3) and Inner Region Round 4 (R4). 

We present only the latest data collected during May and June, 2021, during the afternoon hours with few people (around 20) in the park. In total, we collected 8,353 SigCap data points, with each data-point containing information about all cellular signals received at a particular GPS coordinate. Specifically, there are 44,683 4G, 22,620 LAA/CBRS, and 3,097 5G data points in the measurement set. In addition, we collected 1,333 FCC ST measurements (708 4G, 386 5G and 239 mixed, where the technology changed during the test), with each containing uplink/downlink throughput and latency results. Fig.~\ref{fig:heatmap4G}, \ref{fig:heatmap4Gp}, and \ref{fig:heatmap5G} shows coverage maps of 4G, 4G+LAA/CBRS, and 5G in the park, respectively, from the SigCap measurements.

\begin{figure*}[h!]
\begin{subfigure}[t]{.24\textwidth}
  \centering
    \includegraphics[width=\textwidth]{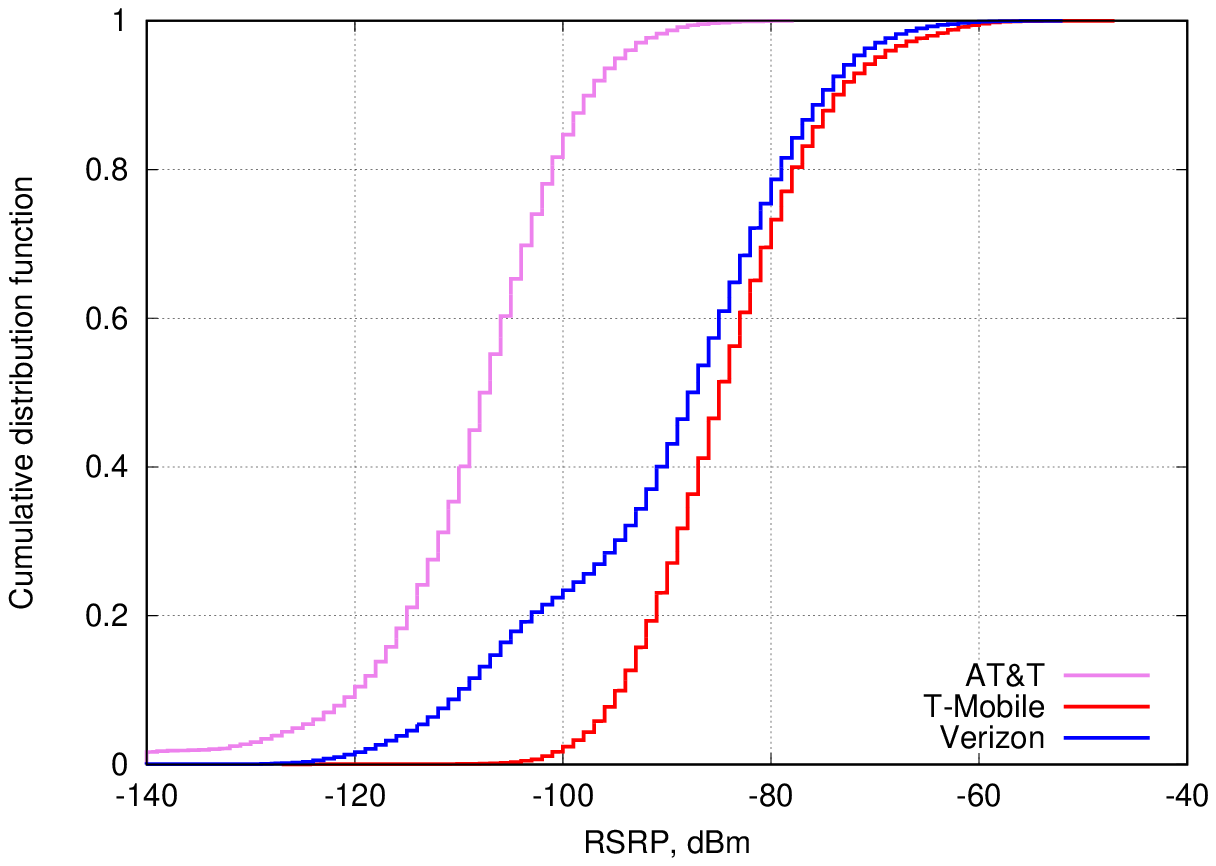}
  \caption{All 4G Channels' RSRP}\label{fig:allRsrp}
\end{subfigure}
\begin{subfigure}[t]{.24\textwidth}
  \centering
  \includegraphics[width=\textwidth]{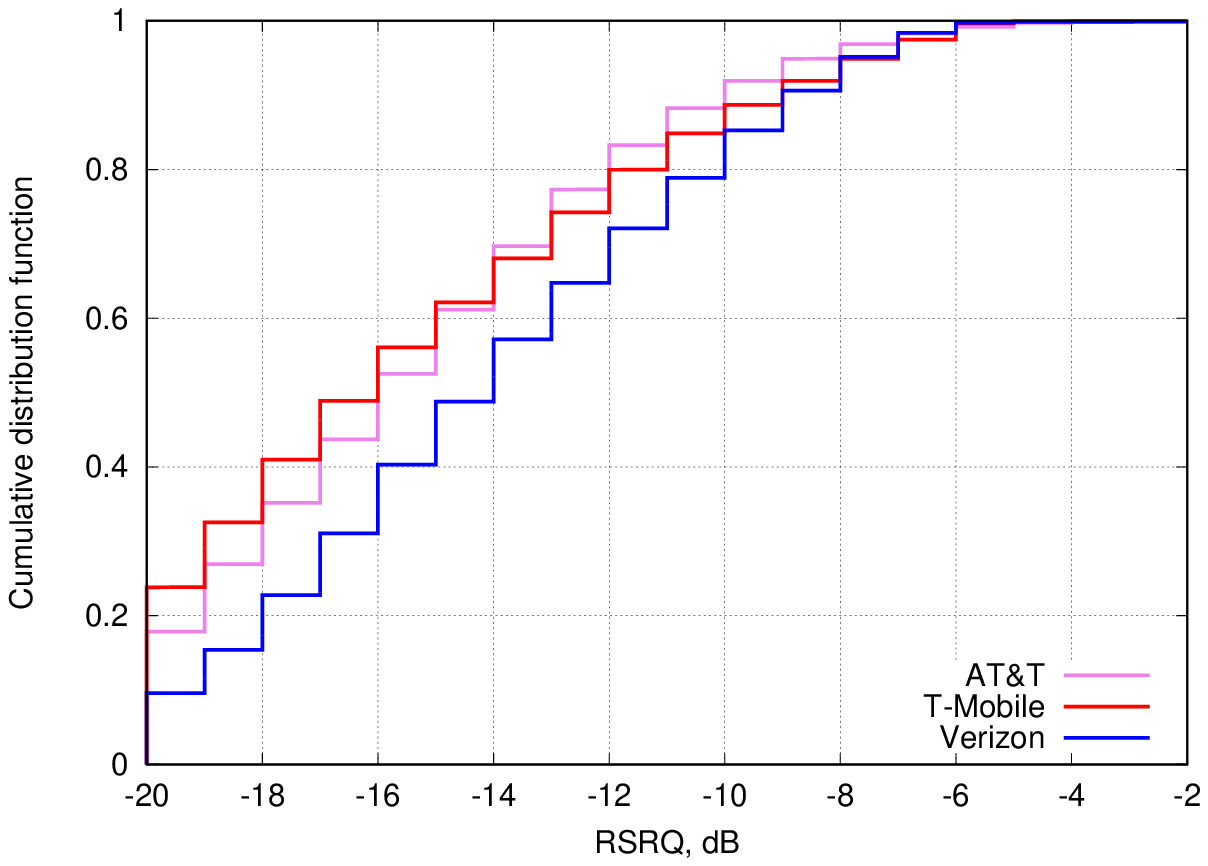}
  \caption{All 4G Channels' RSRQ}\label{fig:allRsrq}
\end{subfigure}
\begin{subfigure}[t]{.24\textwidth}
  \centering
 \includegraphics[width=\textwidth]{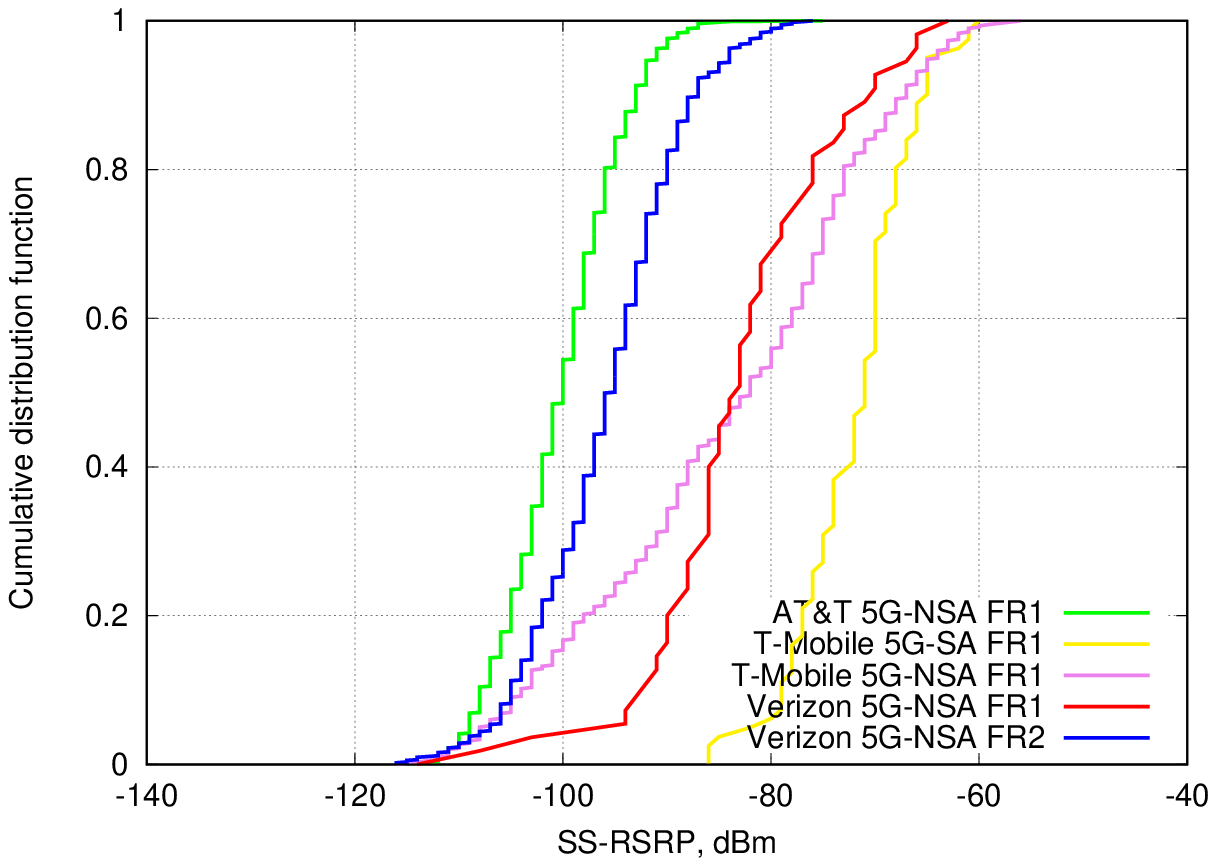}
  \caption{5G RSRP}\label{fig:nrRsrp}
\end{subfigure}
\begin{subfigure}[t]{.24\textwidth}
  \centering
    \includegraphics[width=\textwidth]{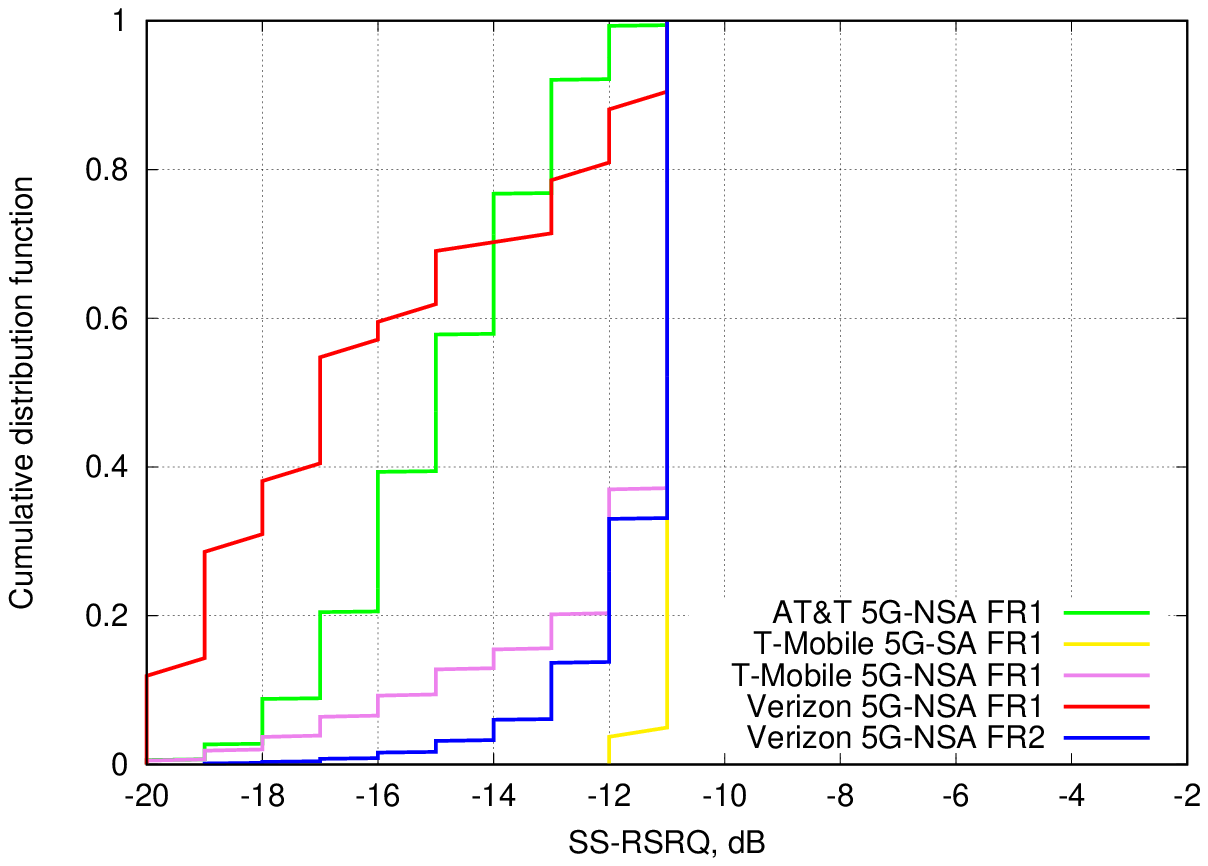}
  \caption{5G RSRQ}\label{fig:nrRsrq}
\end{subfigure}

\caption{AT\&T, T-Mobile and Verizon in Hutchinson Field: CDF of 4G and 5G RSRP, RSRQ}
\label{c}
\end{figure*}

\begin{figure*}[htb!]
\begin{subfigure}{.33\textwidth}
  \centering
 \includegraphics[width=\textwidth]{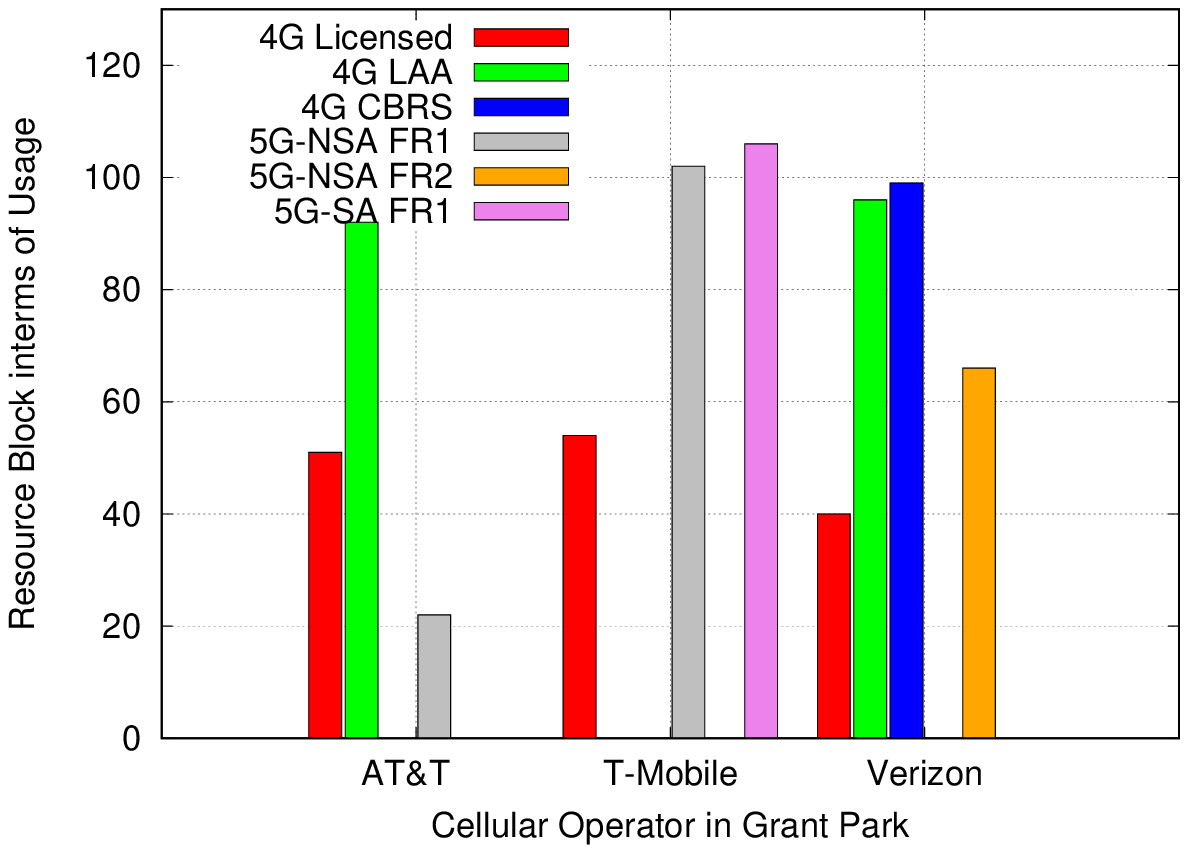}
  \caption{RB Usage}\label{fig:rbUsage}
\end{subfigure}
\begin{subfigure}{.33\textwidth}
  \centering
    \includegraphics[width=\textwidth]{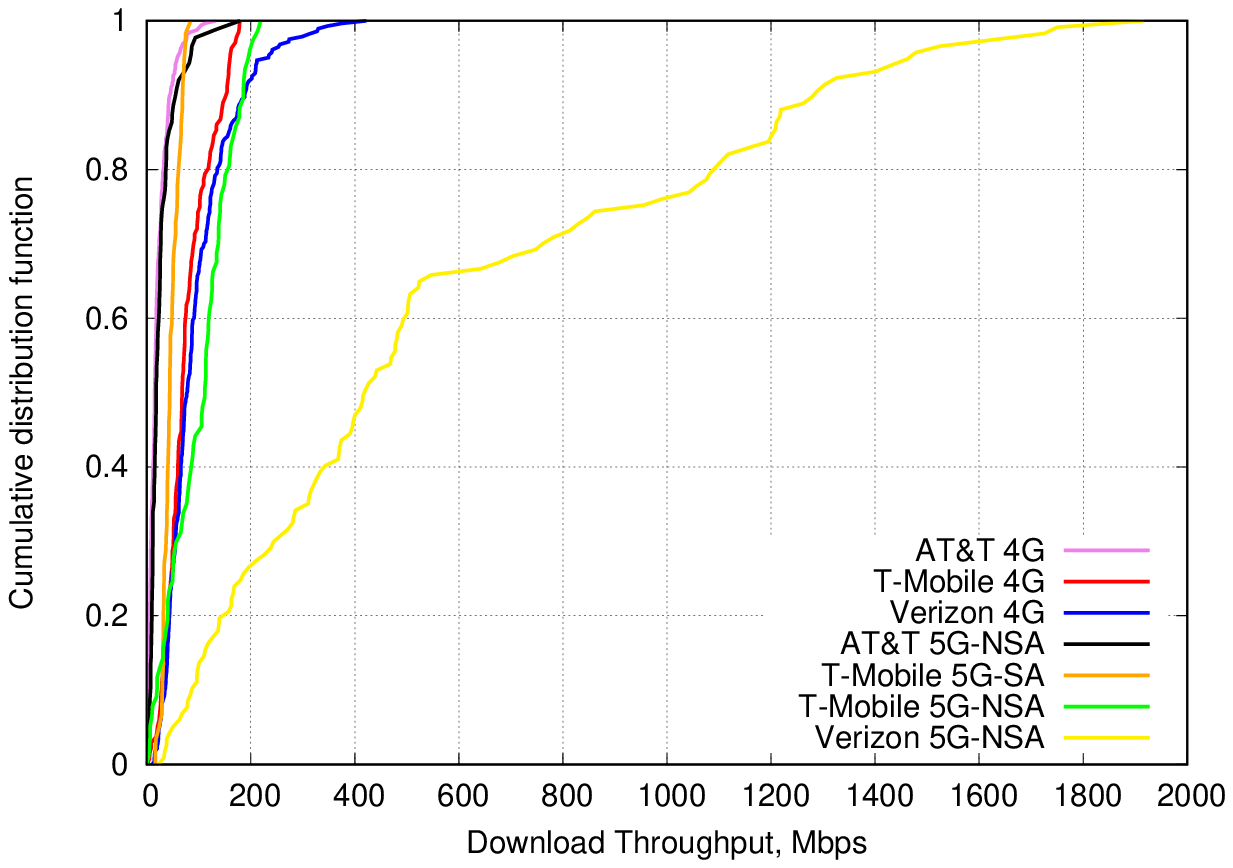}
  \caption{CDF of Downlink Throughput}\label{fig:tput}
\end{subfigure}
\begin{subfigure}{.33\textwidth}
  \centering
   \includegraphics[width=\textwidth]{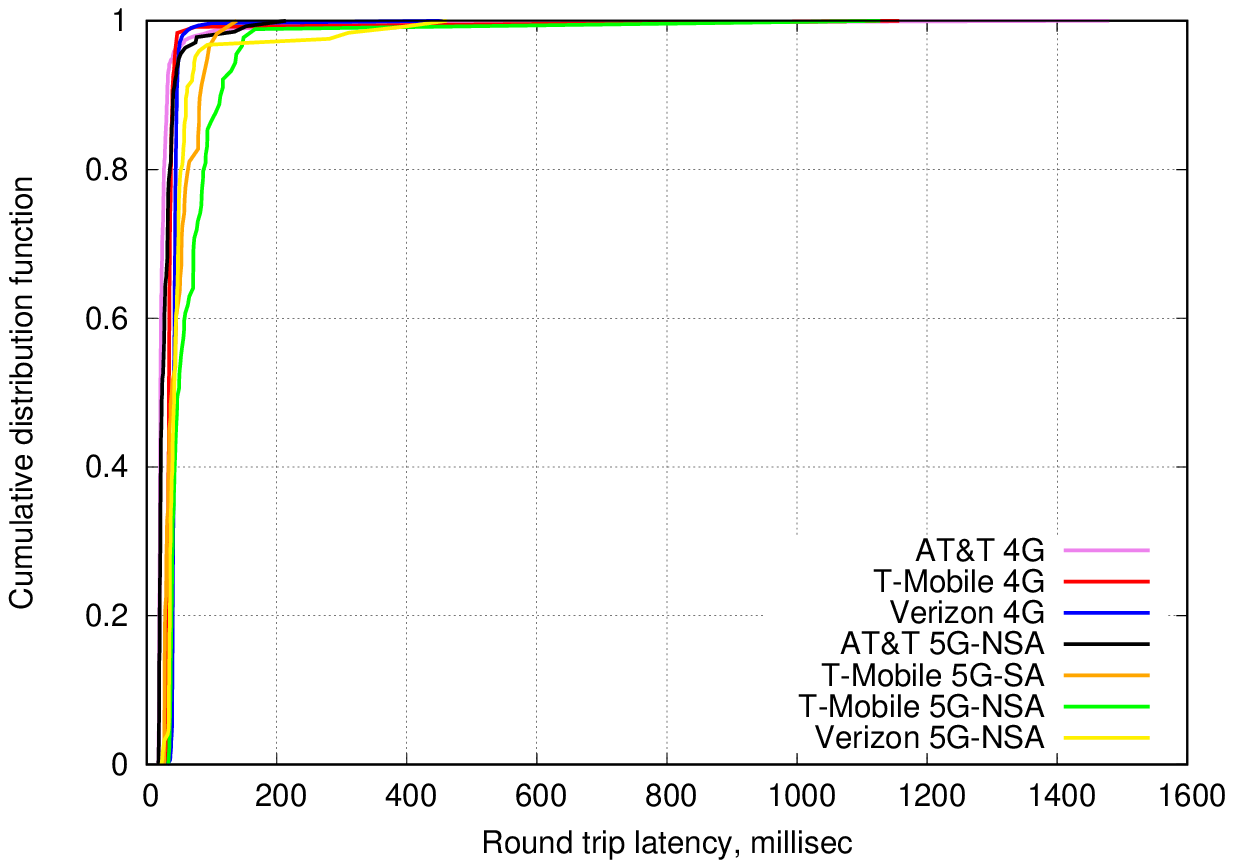}
  \caption{CDF of Round Trip Latency}\label{fig:latency}
\end{subfigure}
\vspace{-0.2cm}
\caption{AT\&T, T-Mobile and Verizon in Hutchinson Field: RB Usage, Downlink Throughput, and Latency}
\label{c}
\end{figure*}

\subsubsection{\textbf{4G deployments in Hutchinson Field}}
Each of the three operators we studied have extensive deployments of 4G in low-band (Bands 5,12,13,14) and mid-band (Bands 2,3,4,\\ 7,30,46,48,66).
We found no AT\&T and T-Mobile BSs deployed inside Hutchinson Field; their 4G bands are mostly deployed on macro-cells located in the greater Grant Park area.
Verizon is the only operator to have deployed 4G and 5G within Hutchinson Field; there are three CBRS (Band 48) channels in 3.56, 3.58, and 3.6 GHz, using General Authorized Access (GAA) \cite{ying2018sas} and LAA (Band 46) channels on two sets of Wi-Fi-equivalent channels: \{36, 40, 44\} in U-NII-1 and \{157, 161, 165\} in U-NII-3.
Additionally, AT\&T has deployed LAA on two sets of channels: \{149, 153, 157\} and \{157, 161, 165\} in U-NII-3.
Both LAA and CBRS were mostly aggregated in groups of three 20 MHz channels with a total bandwidth of 60 MHz, in addition to the primary licensed carrier. 

Channel 157 overlaps the two sets of AT\&T LAA channels and there is also a full overlap between the U-NII-3 channel sets of AT\&T and Verizon, which may lead to a LAA/LAA coexistence problem. Additionally, there is a dense deployment of AT\&T Wi-Fi access points (APs) across the entire 5 GHz unlicensed band in the measurement area. 

\subsubsection{\textbf{5G deployments in Hutchinson Field}}

We identified six lampposts inside the field that are used as Verizon's mmWave BS (blue triangles in Fig.~\ref{fig:gpMap}) using the Ericsson radio.
We found no AT\&T and T-Mobile 5G deployments inside the field.
The average distance between the Verizon mmWave BSs is 140 m (460 ft). Each mmWave antenna panel has a separate PCI  with multiple beam indices. 
Verizon and AT\&T have deployed 5G in NSA mode only, while T-Mobile uses both SA and NSA mode.

T-Mobile and Verizon have deployed 5G low-band in Bands n71 and n5, respectively, using the maximum possible 20 MHz bandwidth, while AT\&T's 5G deployment is in n5 but uses only 5 MHz. These low-band 5G bandwidths are lower than the possible 40 to 100 MHz bandwidths in mid-band. Additionally, due to the limitation of the Pixel 5 being able to aggregate only one 5G channel in FR1, 
the low-band 5G performance is worse than mid-band 4G performance at the present time, since 4G can aggregate up to four channels.

T-Mobile and Verizon have deployed 5G on mid-band and mmWave, respectively as well\footnote{We measured AT\&T 5G mmWave in other areas of downtown Chicago but not in Hutchinson Field as of June 2021.}. T-Mobile's mid-band deployment is in Band n41 using 20 and 80 MHz bandwidths. However, the Pixel 5's limitation of only one secondary 5G carrier in FR1 still applies, leading to a diminished performance compared to 4G at the present time.
On the other hand, Verizon has deployed mmWave 5G densely in n260 (39 GHz) using at most four carriers, each 100 MHz wide. The higher bandwidths and number of channels being aggregated leads to a vastly improved throughput compared to mid-band 5G. Using NSG, we observed that Verizon aggregates mmWave channels only if they were transmitted from the same mmWave panel, i.e. they have the same PCI.

\subsection{Performance Comparison}


\subsubsection{\textbf{Statistical Analysis of RSRP and RSRQ}}

Since SigCap can collect RSRP and RSRQ data for all primary and other channels, we use these values from the SigCap data to create cumulative distribution function (CDF) plots for each operator. 
Fig.~\ref{fig:primaryRsrpBw}, \ref{fig:attEarfcn}, \ref{fig:tmoEarfcn}, and \ref{fig:vzwEarfcn} show the CDF of primary channel RSRP scaled by bandwidth, as an indicator of coverage and throughput performance. The bandwidth scaling is calculated as $RSRP_{dBm} + 10*log_{10}(BW_{MHz})$. We only present the primary channel bandwidth since the API does not provide reliable information on the total bandwidth due to aggregation.

Fig.~\ref{fig:primaryRsrpBw} shows that the BW-scaled RSRP of T-Mobile and Verizon are comparable, while AT\&T's is around 20 dB lower. Similarly, Fig.~\ref{fig:primaryRsrq} shows a higher RSRQ for T-Mobile and Verizon, with AT\&T around 4 dB lower. These CDFs indicate that the 4G performance of T-Mobile and Verizon is better than AT\&T's, which is borne out by throughput analysis presented in the next sub-section.

Next, the CDFs of BW-scaled RSRP sorted by EARFCN are shown, to show each operator's channel selection performance. Fig.~\ref{fig:attEarfcn} shows the CDF for AT\&T, which uses 5 Bands (2,12,14,30,66) as its primary channel, with highest occurrence in bold for Band 2 (EARFCN 675, 57\% of data) and Band 66 (EARFCN 66686, 33\% of data), while Fig.~\ref{fig:attEarfcnRsrq} shows the RSRQ counterparts. From the BW-scaled RSRP we see that Band 14 might be a better choice for the primary channel, however, the difference in RSRP is negligible, and Band 14's RSRQ is around 3 dB lower: since Band 14 is a low-band channel, the propagation is better leading to improved RSRP, but also leads to more neighboring cell interference when the same channel is used on neighboring cells.
Fig.~\ref{fig:tmoEarfcn} and \ref{fig:tmoEarfcnRsrq} show the CDF of BW-scaled primary RSRP and RSRQ for T-Mobile, respectively. There are only two choices for primary channel bands, with Band 66 (EARFCN 66811, 92\% of data) as the majority. This choice seems justified from the RSRP and RSRQ CDFs.
Likewise, Fig.~\ref{fig:vzwEarfcn} shows the CDF of BW-scaled primary RSRP for Verizon. Band 66 (EARFCN 66356, 93\% of data) is selected more often than Band 2 and 13 with higher RSRP. While Fig.~\ref{fig:vzwEarfcnRsrq} shows a lower RSRQ for Band 66 compared to Band 2 and 5.
However, Band 13's RSRQ is slightly lower than Band 66, while Band 2 and 5 has a similar RSRP distribution to Band 66.
The above data indicate that each operator's primary channel choice is based primarily on optimizing RSRP and RSRQ. 
\begin{figure*}[h!]
\begin{subfigure}{.33\textwidth}
  \centering
 \includegraphics[width=0.5\textwidth]{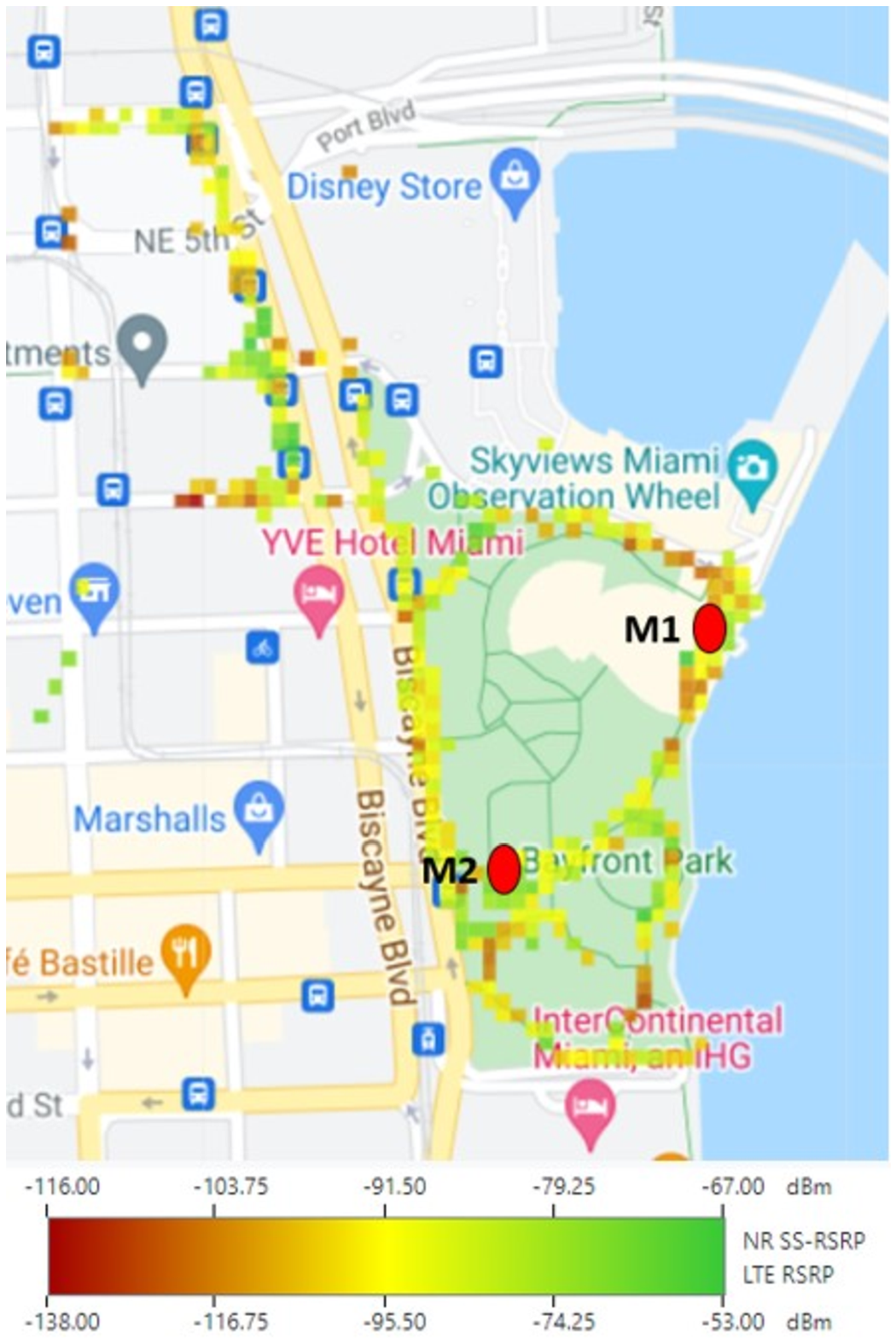}
  \caption{5G Coverage map in Miami downtown }\label{FIU_Miami_Deployment}
\end{subfigure}
~
\begin{subfigure}{.33\textwidth}
  \centering
   \includegraphics[width=\textwidth]{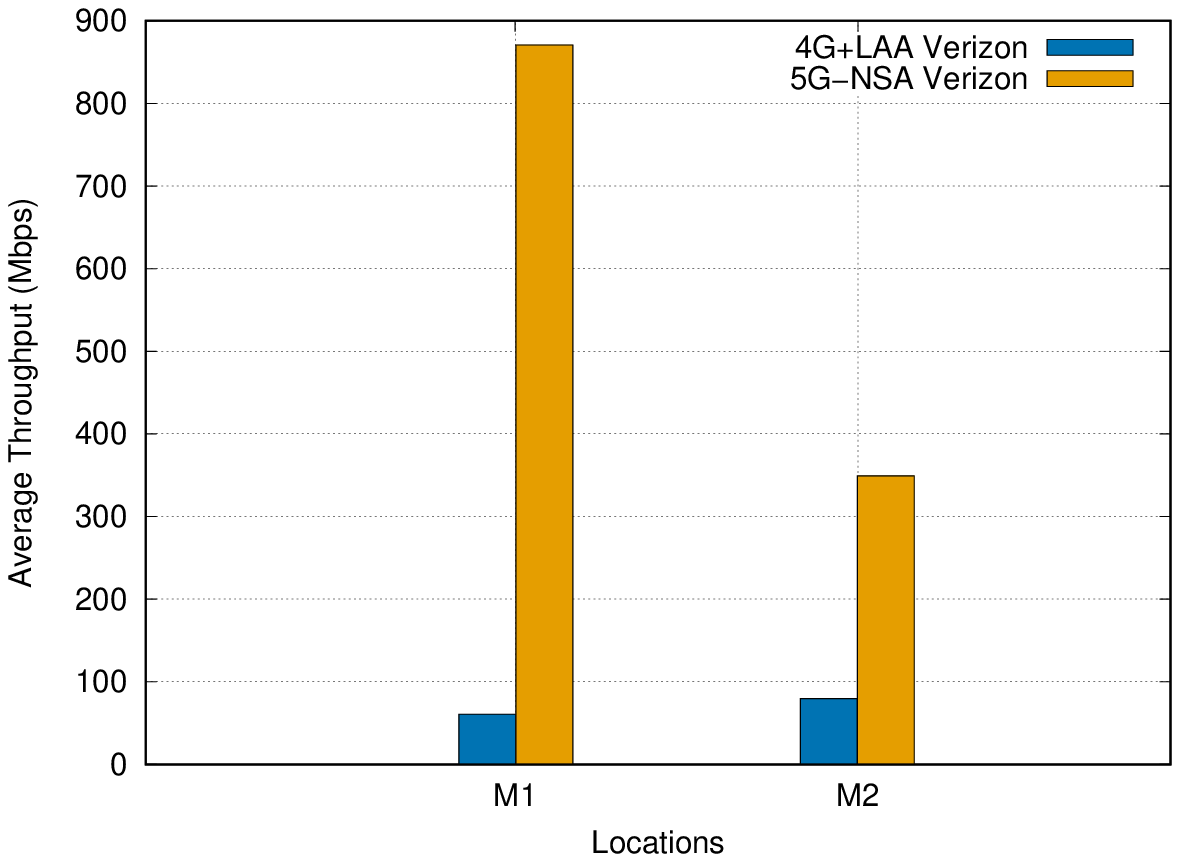}
  \caption{Throughput of 4G+LAA Vs 5G}\label{FIU_4G_5G}
\end{subfigure}
~
\begin{subfigure}{.33\textwidth}
  \centering
   \includegraphics[width=\textwidth]{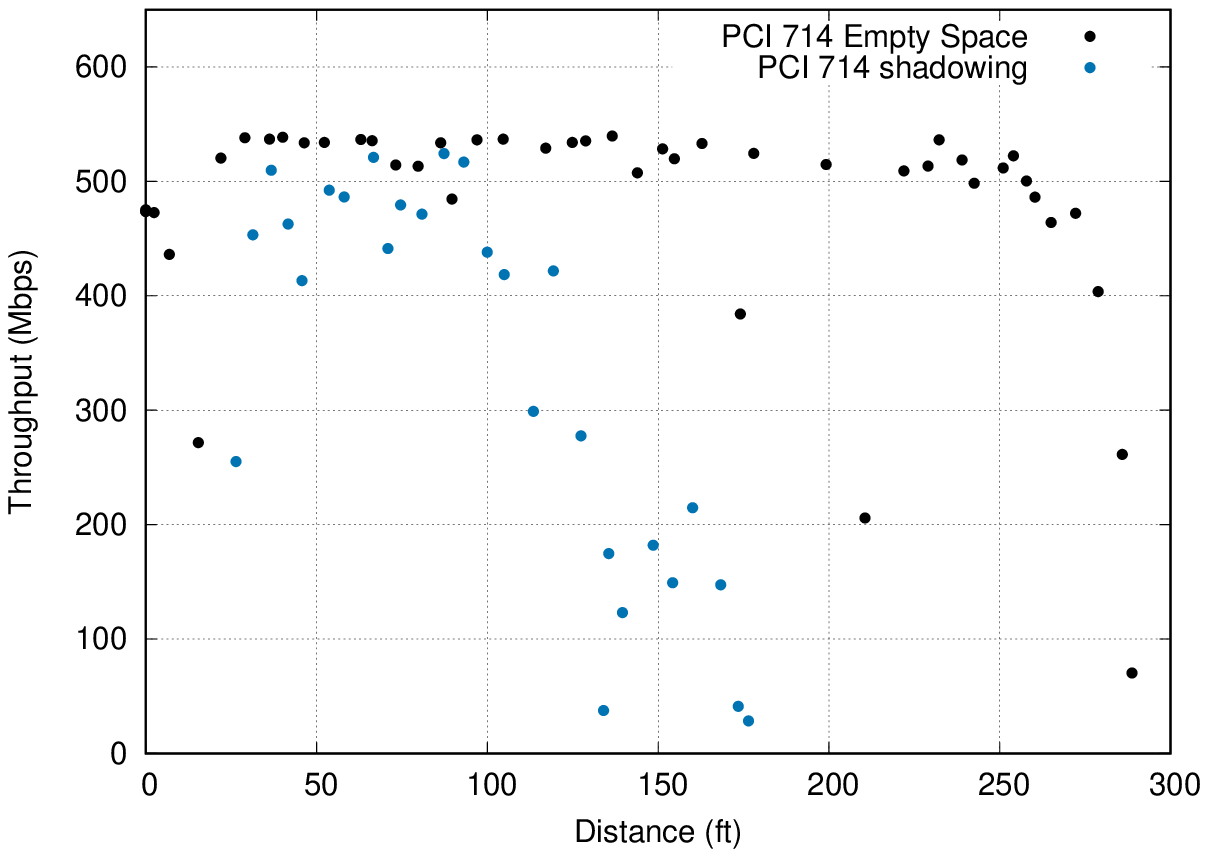}
  \caption{Impact of distance and shadowing}\label{FIU_Distance}
\end{subfigure}
\\
\begin{subfigure}{.33\textwidth}
  \centering
   \includegraphics[width=\textwidth]{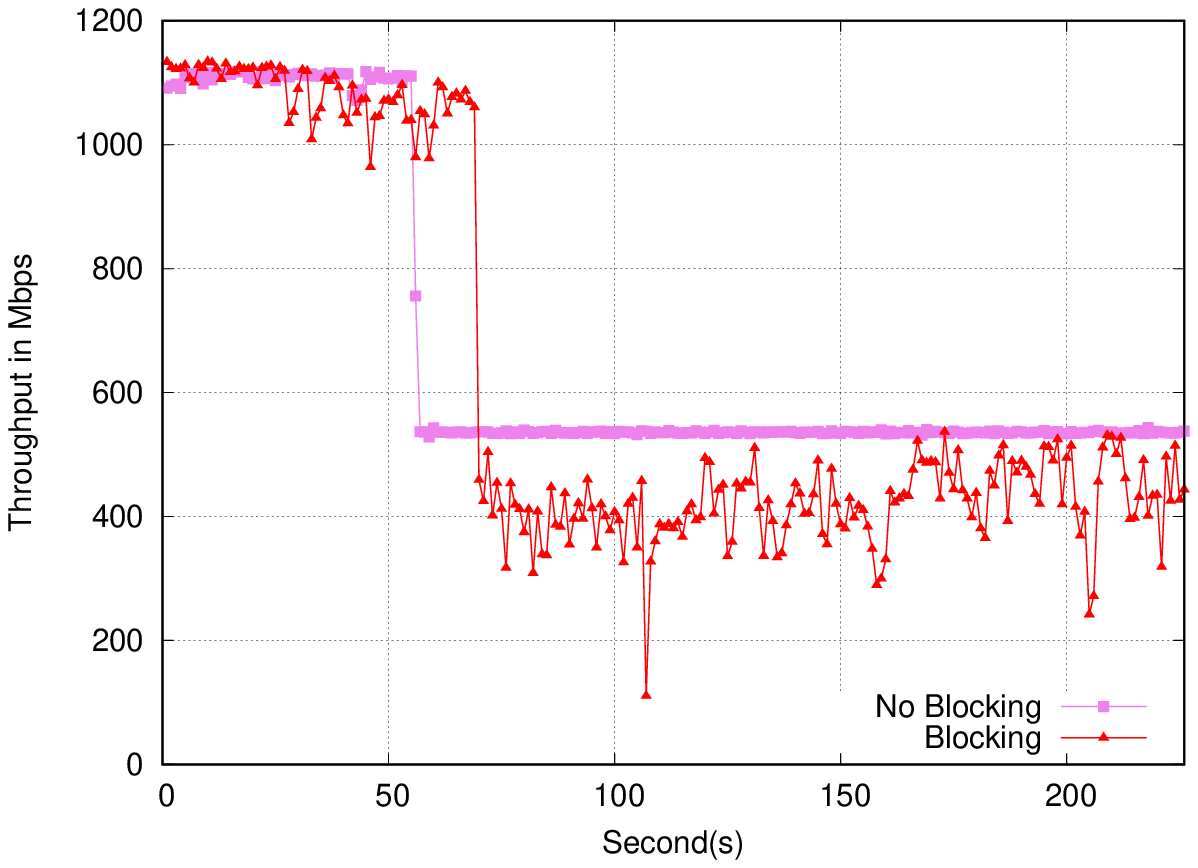}
  \caption{Impact of human body blocking}\label{FIU_Human_Body}
\end{subfigure}
~
\begin{subfigure}{.33\textwidth}
  \centering
   \includegraphics[width=\textwidth]{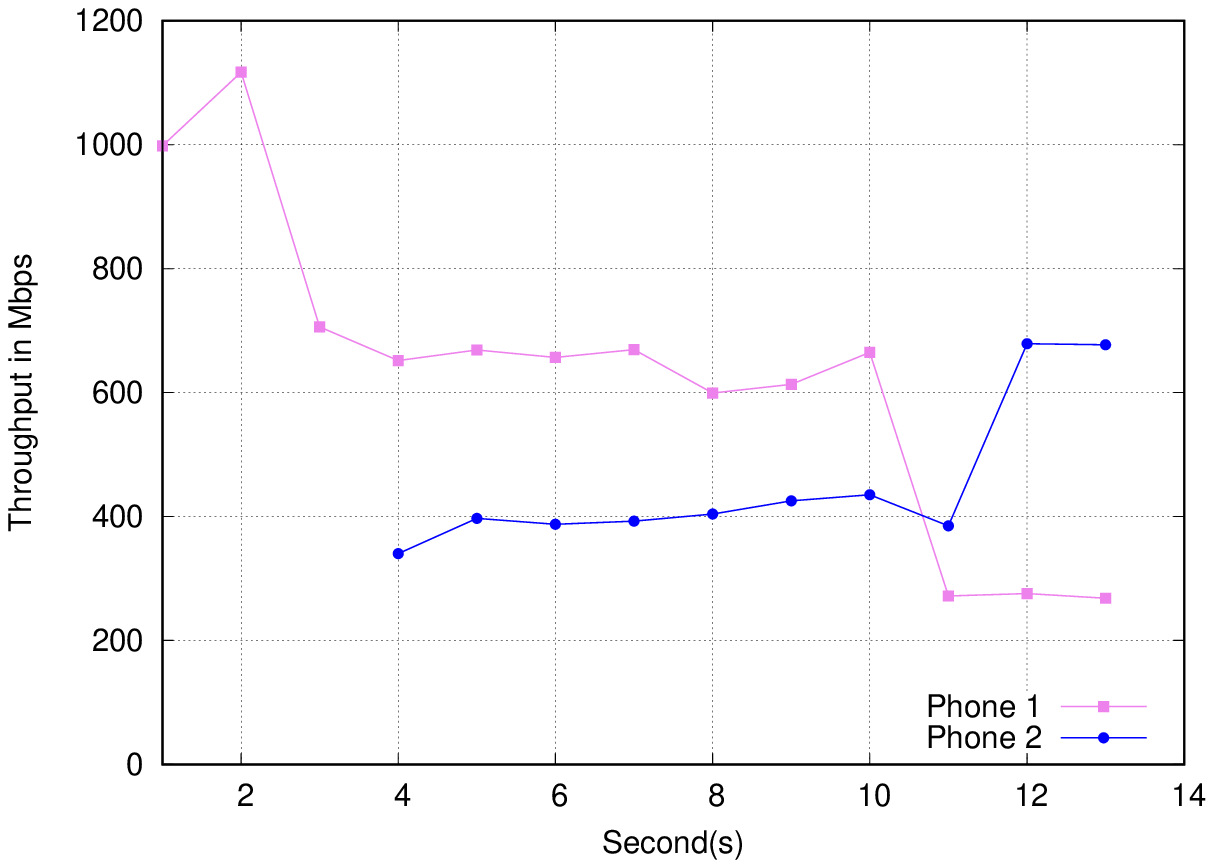}
  \caption{Throughput between 2 phones}\label{FIU_2phones_onePCI}
\end{subfigure}
~
\begin{subfigure}{.33\textwidth}
  \centering
    \includegraphics[width=\textwidth]{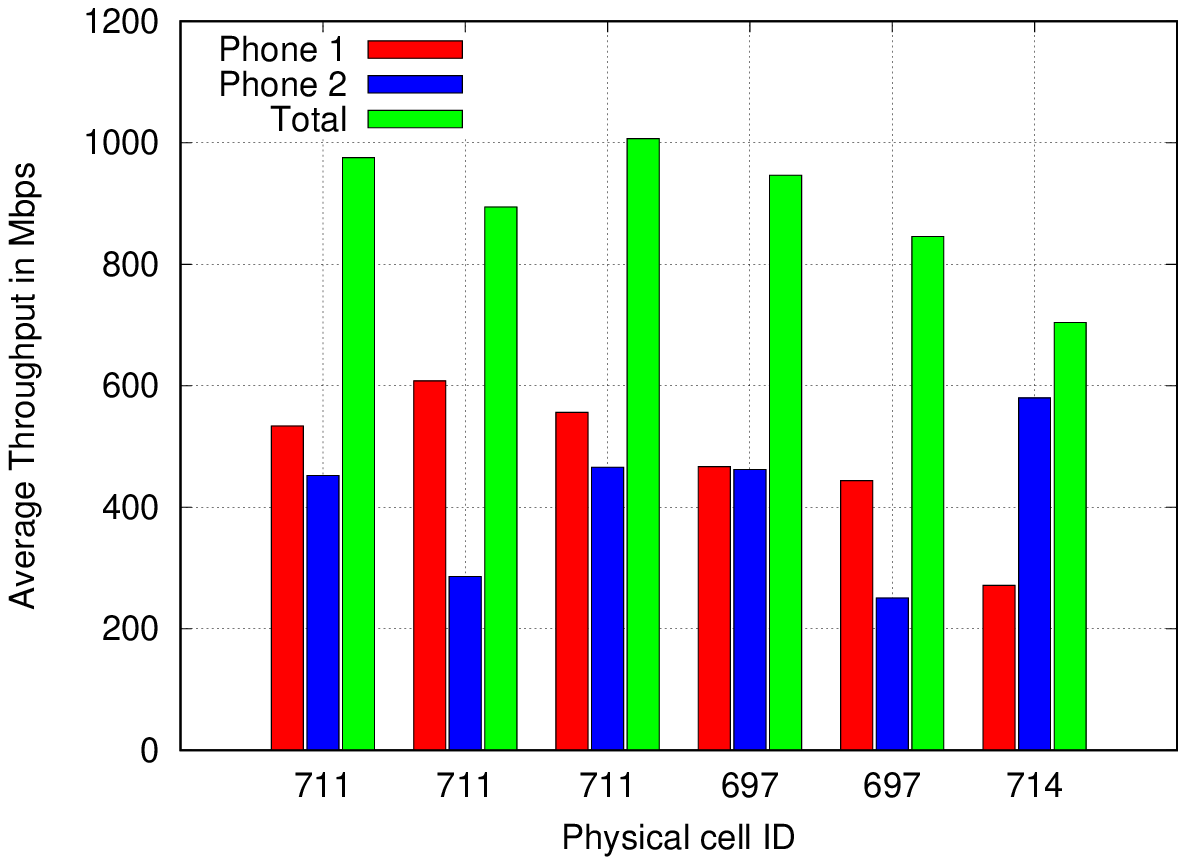}
  \caption{2 phones throughput on multiple PCI}\label{FIU_2phones_multiplePCIs}
\end{subfigure}
\vspace{-0.3cm}
\caption{mmWave Deployment at Miami, Florida}
\label{c}
\end{figure*}

Fig.~\ref{fig:primaryBw} shows the CDF of the primary channel bandwidth. Verizon has the highest available bandwidth for its primary channel, followed by T-Mobile and AT\&T. 
Note that the primary channel bandwidth, RSRP, and RSRQ may not be a good indicator for throughput performance due to carrier aggregation, but does provide insight into the deployment quality: the higher the primary bandwidth and RSRP, the more likely that the operator will have good coverage and throughput. This is corroborated by the throughput analysis in the next sub-section.

Fig.~\ref{fig:allRsrp} and \ref{fig:allRsrq} show the CDF of RSRP and RSRQ for all 4G Licensed carriers (\textit{i.e.}, primary, secondary, neighboring) in Hutchinson Field. Based on this, T-Mobile has the best 4G licensed coverage, followed closely by Verizon and AT\&T. The RSRQ CDF shows Verizon has a better overall channel quality compared to T-Mobile.
On the other hand, AT\&T's RSRP and RSRQ values indicate inferior coverage, which is probably due to the fact that the cells are mostly deployed outside Hutchinson Field.

Fig.~\ref{fig:nrRsrp} and \ref{fig:nrRsrq} show the 5G-RSRP and 5G-RSRQ CDF of 5G when the device is connected to 5G. We do not scale the 5G-RSRP with bandwdith since the app does not provide this information for each data record. Overall, the 5G-RSRP of the FR1 bands is higher than FR2 due to the difference in operating frequency and resultant propagation. The CDF of 5G-RSRP for T-Mobile NSA FR1 deviates from the Gaussian distribution since the values are combined from the low-band (n71) and mid-band (n41), while the 5G-RSRP of T-Mobile SA mode is higher due to the device only connecting to the low-band n71 in SA mode. We were not connected very often to Verizon 5G in FR1 and when we were, the 5G-RSRP and 5G-RSRQ values were generally lower. When the device was blocked from connecting to 5G mmWave (using NSG's root access), the device would connect more often to 4G+LAA/CBRS rather than mid-band 5G, perhaps because the former configuration provided higher throughput. Finally, we observed a very low 5G-RSRP and 5G-RSRQ of AT\&T FR1, indicating inferior 5G coverage.

While LAA and CBRS information was collected, we do not include them in the comparisons since there is a substantial difference in transmit power compared to the licensed channels; the U-NII-3 spectrum used by LAA, allows a maximum of 30 dBm transmit power, while CBRS allows a maximum of 47 dBm in outdoor deployments.

From NSG, we show the average RB allocation per device as an indicator of network load in Fig.~\ref{fig:rbUsage}. 
There are slightly fewer RBs allocated on Verizon's licensed carrier compared to the other operators, indicating a higher load or higher resource allocations on the secondary LAA/CBRS/5G carriers. However, the difference is not significant enough, and we can conclude that the network load is similar for all operators during the measurements.

\subsubsection{\textbf{Downlink Throughput and Latency Performance, using FCC ST}}

The data was sorted based on the cellular technology used: we removed data where the technology switched between 4G and 5G during the test. The SA tests (only on T-Mobile) were run by forcing the phone to use SA only using root access.

Fig.~\ref{fig:tput} shows the downlink throughput CDF of AT\&T, T-Mobile, and Verizon in 4G and 5G. AT\&T had the worst 4G and 5G throughput in Hutchinson Field due to low coverage and low bandwidth (5 MHz) of Band n5.
Verizon 5G mmWave had the best throughput:  
the maximum throughput achieved was 1.92 Gbps, which is constrained by Pixel 5's support of a maximum of four aggregated mmWave channels \footnote{Other 5G phones may have higher maximum downlink throughput due to greater mmWave aggregation capability, \textit{e.g.}, Samsung Galaxy S21 Ultra supports a maximum of eight aggregated mmWave channels.}. Most of the FCC ST data for Verizon 5G was captured using mmWave since there was a sparse deployment of 5G in FR1. 

The next best throughput performance is achieved closely by Verizon 4G and T-Mobile in 4G and 5G-NSA. Both Verizon and T-Mobile achieved a very similar performance in 4G, which correlates to the similarity of their 4G RSRP, RSRQ, and primary bandwidth distribution. However, Verizon delivered the highest 4G throughput of 421 Mbps due to LAA/\\CBRS usage, which is better than the highest 5G throughput in FR1 of 219 Mbps, achieved by T-Mobile 5G-NSA.
Due to device limitations, only a maximum of one secondary 5G FR1 carrier can be aggregated. Thus, there is a diminished throughput increase in T-Mobile between 5G-NSA and 4G, even though 80 MHz is available on T-Mobile's 5G channel in Band n41. Similarly, T-Mobile 5G-SA offered low throughputs due to the single 5G channel usage, without even a 4G primary channel.
The average download throughput recorded in the Hutchinson Field region for all operators are as follows: (i) AT\&T: 20.7 Mbps and 27.1 Mbps in 4G and 5G-NSA, respectively; (ii) T-Mobile: 77.2 Mbps, 46.2 Mbps, and 101.3 Mbps in 4G, 5G-SA, and 5G-NSA, respectively and (iii) Verizon: 95.8 Mbps and 574.4 Mbps in 4G and 5G-NSA, respectively. Verizon achieved the best throughput performance due to its usage of mmWave. 

Fig.~\ref{fig:latency} shows the CDF of the round trip idle latency of the three operators over 4G and 5G. The median values are: 30.5 ms and 30.7 for AT\&T 4G and 5G-NSA, respectively; 44.1 ms, 48.4 ms, and 74.8 ms for T-Mobile 4G, 5G-SA and 5G-NSA, respectively; 44.1 ms and 54.4 ms for Verizon 4G and 5G-NSA, respectively. Generally, the latency performance is poorer in 5G-NSA compared to 4G. This may be due to non-optimal deployment of 5G-NSA, causing additional overheads due to dual connectivity. It should be noted that the latency measurement is end-to-end, however, since all the latency tests were conducted via the same two servers, the effects of back-haul on the latency are the same for all the operators. We did not notice any significant difference in throughput and latency between tests conducted over the two servers. 

It is clear that 5G mmWave provides a significantly improved throughput performance, but the latency performance could be improved. In spite of the directional nature of mmWave transmissions, the dense deployment of 6 BSs over 0.1 km$^2$, with average distance of 140 m between BSs provides very good 5G mmWave coverage in Hutchinson Field. However, the directional nature also results in a higher variance of 5G mmWave throughput as seen in Fig.~\ref{fig:tput}. Hence, in the next section, we focus on a single Verizon 5G mmWave BS to better quantify mmWave performance as a function of distance, body loss, and number of clients.
\vspace{-0.2cm}
\section{Measurements in Miami}

We utilized two Pixel 5 phones as summarized in Table~\ref{devices}. We collected 4G, 4G+LAA, and 5G measurements on only the Verizon network while walking in the park and city streets within the downtown area shown in Fig.~\ref{FIU_Miami_Deployment}. The Miami measurements were done between January and June 2021.
Verizon has a diverse deployment in downtown Miami with a mix of 4G, 4G+LAA, and 5G mmWave, as was previously summarized in Table~\ref{opd}. Unlike Hutchinson Field in Chicago where CBRS has been widely deployed, CBRS was not detected in Miami, presumably due to the coastal location with more radar deployment. The Verizon mmWave operating band is n261 (28 GHz) unlike Chicago where it was n260 (39 GHz) with a bandwidth of 400 MHz (aggregated over four carries, each 100 MHz).

First, Fig.~\ref{FIU_Miami_Deployment} shows the \emph{coverage map} of 5G deployment in downtown Miami. 4G+LAA is also widely deployed in the same area and Fig.~\ref{FIU_4G_5G} shows the advantage in throughput of 5G compared to 4G+LAA at two different locations (M1 and M2), as indicated in Fig.~\ref{FIU_Miami_Deployment}. The 5G throughput gain is in the range of $4\times$ to $14\times$, compared to 4G+LAA. Second, Fig.~\ref{FIU_Distance} shows the impact of \emph{distance} on the 5G mmWave coverage. As shown, the maximum throughput is achieved up to $250$ feet before it dramatically drops down at $300$~ft. Furthermore, having trees (\textit{i.e.}, shadowing effect) reduces the coverage range down to $125$~ft (\textit{i.e.}, $50\%$ drop in coverage). 
Third, Fig.~\ref{FIU_Human_Body} depicts the impact of \emph{human body blocking}, in which $2$ different trials were conducted. One trial had the user's body blocking the phone, while the other did not. The trials were conducted at a fixed distance to the tower with no other obstructions, both phones were connected to the same PCI $714$, and the same beam number throughput the trial. The average degradation in throughput due to human body blockage is about $20\%$.
 
Finally, Fig.~\ref{FIU_2phones_onePCI} shows the impact of having \emph{two simultaneously} served phones. In this experiment, we use two Google Pixel 5 phones. 
The two phones were held within arms-length of one another near a cell tower, and phone-2 starts four seconds after phone-1. Fig.~\ref{FIU_2phones_onePCI} depicts the throughput achieved by each of the two phones over time. As shown, phone-1 starts with a high throughput, which is an indication that it is being allocated all the available resource blocks. Once phone-2 starts, the throughput of phone-1 drops given that the total resource blocks are now shared between the two phones. Such an experiment was repeated multiple times, and Fig.~\ref{FIU_2phones_multiplePCIs} shows the outcome of $6$ such trials over different PCIs. In most cases, the throughput values of both phones are comparable. 

\section{Conclusions and future work}

The methodology developed in this paper, using a variety of apps on smartphones, is a quick, scalable, way of obtaining comprehensive information about complex cellular deployments that use a mix of frequency bands and technologies. At the present time, 5G deployments are evolving rapidly and such measurement campaigns enable researchers to uncover issues that can be further studied on experimental test-beds. It is clear that 5G performance will continue to improve, both in network deployment as well as device performance. Some of the research issues uncovered by the work presented in this paper, which we plan to address in future research, are: (i) it appears from our measurements that operator's choice of primary channel is primarily determined by RSRP and RSRQ. It is not clear however if this choice correlates with higher throughput. We would like to explore learning algorithms based on the data we have collected to determine if there are better channel choices, given the increasingly large number of channel aggregation options available to operators; (ii) the latency performance of 5G is worse than 4G at present. This is most likely due to 5G being deployed in NSA mode. However, even the limited SA data available on T-Mobile's mid-band 5G network does not exhibit improved latency. We would like to focus on this aspect in our future work and include latency under load measurements (FCC ST only measures idle latency) ; (iii) 4G with aggregated channels in the unlicensed LAA and CBRS bands can deliver throughput in the mid-band that is comparable or even higher than that offered by mid-band 5G: Verizon's maximum 4G throughput, using LAA and CBRS is 421 Mbps compared to T-Mobile's 5G throughput of 219 Mbps. However, as more LAA and CBRS deployments roll out there will be coexistence issues that will need to be addressed with LAA and synchronization uses for the TDD deployment in CBRS; and (iv) from our in-depth study of 5G mmWave in Miami, we observed that even though mmWave has significantly higher data rate compared to 4G+LAA/CBRS, this higher performance cannot be guaranteed in all locations, due to distance limitations, body loss, and non-line-of-sight to the mmWave BS caused by foliage and other obstructions. To maintain a reliable connection for future applications like AR/VR, having wider and more robust coverage via mid-band 4G and/or 5G seems essential. Reducing the variance of 5G mmWave throughput will be a focus of our future work. 


\end{document}